\UseRawInputEncoding
\documentclass[11pt,twocolumn]{article}
\usepackage[ margin=1in]{geometry}
\usepackage[affil-it]{authblk}
\usepackage{setspace}

\usepackage[sort&compress,numbers]{natbib}
\usepackage[font=small,justification=justified,format=plain]{caption}
\usepackage{graphicx}
\usepackage{amsmath,amssymb,bm}
\usepackage{xr}
\usepackage{color}
\usepackage[normalem]{ulem}
\usepackage{wasysym}
\usepackage{siunitx}
\usepackage{physics}
\usepackage{gensymb}
\usepackage[T1]{fontenc}
\usepackage{booktabs}
\usepackage[section]{placeins}
\usepackage{textcomp}
\usepackage[colorlinks=true, linkcolor=black, urlcolor=blue, citecolor=blue]{hyperref}
\usepackage{verbatim}

\makeatletter

\newcommand{\moire}{moir\'e }
\newcommand{\Moire}{Moir\'e }
\newcommand{\Vxxw}{$V_{xx}^\omega$}
\newcommand{\Vxytw}{$V_{xy}^{2\omega}$}
\newcommand{\Vbynm}{V\,nm$^{-1}$}

\makeatother

\def\be{\begin{equation}}
	\def\ee{\end{equation}}
\def \bea{\begin{eqnarray}}
	\def \eea{\end{eqnarray}}
\def \nn{\nonumber}

\begin{document}
	\title{Berry curvature dipole senses topological transition in a \moire superlattice}
	\author[1$\dagger$*]{Subhajit Sinha}
	\author[1$\dagger$]{Pratap Chandra Adak}
	\affil[1]{Department of Condensed Matter Physics and Materials Science, Tata Institute of Fundamental Research, Homi Bhabha Road, Mumbai 400005, India.}
	
	\author[2]{Atasi Chakraborty}
	\affil[2]{Department of Physics, Indian Institute of Technology, Kanpur 208016, India.}
	\author[2]{Kamal Das}
	
	\author[3]{Koyendrila Debnath}
	\affil[3]{Theoretical Sciences Unit, Jawaharlal Nehru Centre for Advanced Scientific Research, Bangalore 560064, India.}
	
	\author[1]{L. D. Varma Sangani}
	
	\author[4]{Kenji Watanabe}
	\affil[4]{Research Center for Functional Materials,
		National Institute for Materials Science, 1-1 Namiki, Tsukuba 305-0044, Japan.}
	
	\author[5]{Takashi Taniguchi}
	\affil[5]{International Center for Materials Nanoarchitectonics,
		National Institute for Materials Science,  1-1 Namiki, Tsukuba 305-0044, Japan.}
	
	\author[3]{Umesh V. Waghmare}
	
	\author[2*]{Amit Agarwal}
	\author[1*]{Mandar M. Deshmukh}
	
	\affil[$\dagger$]{\textnormal{These two authors contributed equally to this work}}
	\affil[*]{\textnormal{sinhasubhajit25@gmail.com, amitag@iitk.ac.in, deshmukh@tifr.res.in}}
	
	\date{}
	\maketitle
	\newpage
	\setlength{\columnsep}{5cm}
	
	\textbf{Topological aspects of electron wavefunction play a crucial role in determining the physical properties of materials.
		Berry curvature and Chern number are used to define the topological structure of electronic bands.
		While Berry curvature and its effects in materials have been studied~\cite{xiao_berry_2010-2,nagaosa_anomalous_2010-1}, detecting changes in the topological invariant, Chern number, is challenging.
		In this regard, twisted double bilayer graphene (TDBG)~\cite{burg_correlated_2019,shen_correlated_2020,adak_tunable_2020_prb,cao_tunable_2020,liu_tunable_2020-1} has emerged as a promising platform to gain electrical control over the Berry curvature hotspots~\cite{sinha_bulk_2020-4} and the valley Chern numbers of its flat bands~\cite{koshino_band_2019-1,zhang_nearly_2019}.
		In addition, strain induced breaking of the three-fold rotation ($C_3$) symmetry in TDBG, leads to a non-zero first moment of Berry curvature called the Berry curvature dipole (BCD), which can be sensed using nonlinear Hall (NLH) effect~\cite{fu_quantum_2015}.
		We reveal, using TDBG, that the BCD detects topological transitions in the bands and changes its sign. 
		In TDBG, the perpendicular electric field tunes the valley Chern number and the BCD simultaneously allowing us a tunable system to probe the physics of topological transitions. 
		Furthermore, we find hysteresis of longitudinal and NLH responses with electric field that can be attributed to switching of electric polarization in \moire systems.
		Such a hysteretic response holds promise for next-generation Berry curvature-based memory devices.
		Probing topological transitions, as we show, can be emulated in other 3D topological systems.}
	
	Exploring materials with new topological phases and probing their symmetries has been at the forefront of modern research.
	Topology is often characterized by Berry curvature that manifests as quantum Hall effect~\cite{xiao_berry_2010-2} or as anomalous Hall effect in magnetic materials~\cite{xiao_berry_2010-2, nagaosa_anomalous_2010-1}.
	While these effects have led to many breakthroughs in physics, these linear responses vanish in systems that preserve time-reversal symmetry. 
	However, even in time-reversal symmetry protected systems, broken inversion symmetry can lead to the first moment of Berry curvature, namely Berry curvature dipole (BCD), when further spatial symmetries are reduced~\cite{fu_quantum_2015}. 
	In presence of an ac current (of frequency $\omega$) in such system, the BCD can drive a second-order electrical response, namely the nonlinear Hall (NLH) voltage with zero  (DC) and second-harmonic ($2\omega$) frequency.
	Aligned with the recent interest in various nonlinear phenomena such as nonlinear optics, there is a rapidly growing interest to look for quantum materials that host the NLH effect~\cite{ma_topology_2021,du_nonlinear_2021}.

	The topological phase of quantum materials can be characterized by the band-specific topological invariant Chern number.
	Often systems undergo transition between topological phases and these transitions are hard to detect, unlike phase transitions of order parameter.
	Recent proposals suggest that topological transitions are accompanied by simultaneous changes in BCD~\cite{facio_strongly_2018,hu_nonlinear_2020}. 
	\Moire systems are known to be natural candidates to host topological bands~\cite{song_topological_2015}.
	While the NLH effect has been observed for transition metal dichalcogenides (TMDCs)~\cite{kang_nonlinear_2019,ma_observation_2019,shvetsov_nonlinear_2019,xiao_berry_2020,tiwari_giant_2021} and corrugated bilayer graphene~\cite{ho_hall_2021}, the experimental study in two-dimensional \moire systems is limited~\cite{huang_giant_2021}.
	Furthermore, topological transitions along with BCD have not been experimentally observed thus far.

	Recently, the observation of several novel phenomena in \moire systems such as magic-angle twisted bilayer graphene has attracted attention~\cite{kim_tunable_2017-2, cao_correlated_2018, cao_unconventional_2018, lu_superconductors_2019, sharpe_emergent_2019-1, serlin_intrinsic_2020}. 
	Flat bands in twistronic systems  give rise to various electron correlation phenomena such as Mott insulators and superconductivity~\cite{kim_tunable_2017-2,cao_correlated_2018,cao_unconventional_2018,lu_superconductors_2019}. 
	In particular, the observation of anomalous Hall effect and orbital ferromagnetism point to the rich topology of these twisted systems~\cite{sharpe_emergent_2019-1,serlin_intrinsic_2020}. 
	Flat bands of \moire systems are also susceptible to  symmetry breaking by strain~\cite{zhang_giant_2020}. 
	Together with the fact that spatial symmetry breaking can lead to nonzero BCD~\cite{zhang_giant_2020, pantaleon_tunable_2021,he_giant_2021}, bands in 2D \moire systems offer an interesting platform to electrically tune topology and detect it via the NLH effect.

	\begin{figure*}
		\centering
		\includegraphics[width=15.5cm]{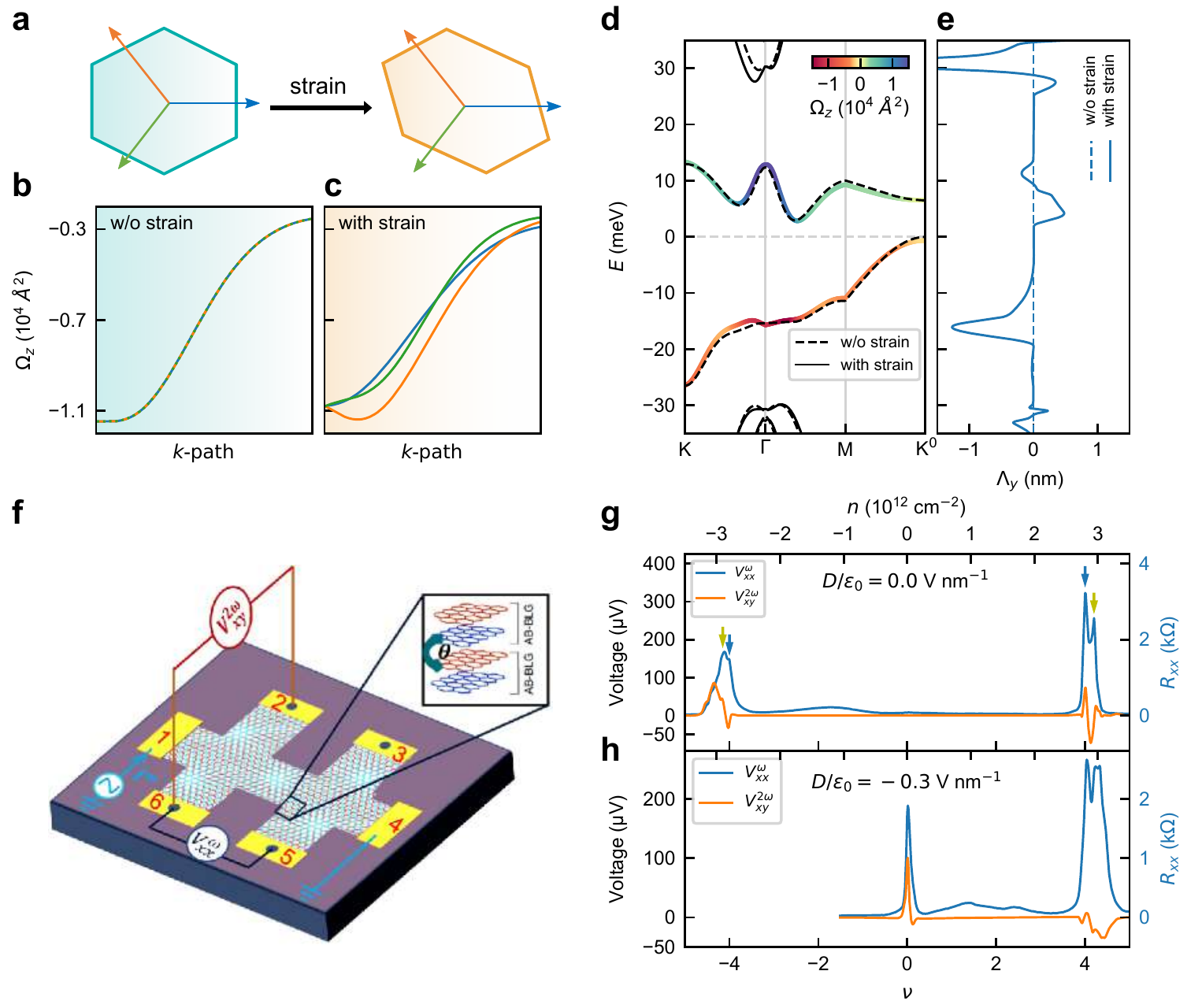}
		\caption{ \label{fig:fig1} { \textbf{Strain-mediated nonlinear Hall (NLH) effect in twisted double bilayer graphene (TDBG).} 
				\textbf{a,}~Schematic shows the distortion of the hexagonal \moire Brillouin zone due to strain, that breaks the $C_3$ symmetry.
				\textbf{b, c,}~The Berry curvature ($\Omega_z$) of the valence band  without strain (\textbf{b}) and with strain (\textbf{c}) along three paths arranged in 120$^\circ$ angle as shown in \textbf{a}.
				\textbf{d,}~The band structure of 1.10$^\circ$ TDBG device with (solid lines) and without (dashed lines) 0.1$\%$ strain for an inter-layer potential of 11 meV, which opens up a gap between the low-energy flat bands along with finite \moire gaps. The energy is measured from the Fermi energy for zero doping. The overlayed color indicates the Berry curvature of the flat bands. 
				\textbf{e,}~The corresponding variation of the $y$-component of the Berry curvature dipole ($\Lambda_y$) for the 0.1$\%$ strained (solid line) and unstrained (dashed line) cases at 2~K. In the absence of strain there is no Berry curvature dipole due to the presence of the $C_3$ symmetry.
				\textbf{f,}~Schematic of the NLH voltage measurement scheme without the encapsulating hBNs and top metal gate for clarity. When a current~($I$) with frequency $\omega$ is sent from 1-4, the voltage drop between 6-5 at frequency $\omega$ gives the longitudinal voltage \Vxxw{} and the voltage drop between 2-6 at frequency 2$\omega$ gives the NLH voltage \Vxytw. The schematic of TDBG in the inset indicates the two Bernal stacked AB bilayer graphenes, with an intermediate twist angle of $\theta = 1.10^\circ$. 
				\textbf{g, h,}~NLH voltage (\Vxytw) (orange) and longitudinal voltage (\Vxxw) (blue) vs. filling factor ($\nu$) at $D/\epsilon_0=0$~\Vbynm{} (\textbf{g}) and $D/\epsilon_0=-0.3$~\Vbynm{} (\textbf{h}). The data is taken at 1.5~K with a current of 100~nA. The right axes indicate the corresponding longitudinal resistance, $R_{xx}$. The two peaks in \Vxxw{} at $\nu=\pm4$, marked by blue and green arrows in \textbf{g}, indicate \moire peaks due to an angle inhomogeneity of $\sim0.03^\circ$, a signature of strain. The data in \textbf{h} is truncated towards negative $\nu$ due to restrictions in gate voltage that can be applied without causing a dielectric breakdown of the hBN.}}
	\end{figure*}

	In this work, we use twisted double bilayer graphene (TDBG), in which two copies of bilayer graphene are stacked together with a small twist angle $\sim$1.1$^\circ$, as a candidate system to study topological transitions using the NLH effect. 
	While the flat bands in TDBG host correlation induced physics as in twisted bilayer graphene, TDBG is additionally equipped with electric field tunability~\cite{burg_correlated_2019, shen_correlated_2020, adak_tunable_2020_prb,cao_tunable_2020, liu_tunable_2020-1, he_symmetry_2020}.
	The perpendicular electric field, apart from modulating the band structure, can also change the valley Chern number of a flat band~\cite{zhang_nearly_2019}.
	Additionally, almost touching flat bands with small tunable energy gaps of few meV in TDBG gives rise to large Berry curvature in this system~\cite{song_topological_2015,chebrolu_flat_2019,sinha_bulk_2020-4}.
	We use electric field tunability to demonstrate that TDBG hosts substantially large BCD due to strain.
	The BCD changes sign abruptly and this can be attributed to a topological transition, a change in valley Chern number, with the electric field.
	Additionally, hysteresis in both longitudinal resistance and NLH signal as a function of the electric field suggests switching between metastable states.
	
	In TDBG, broken inversion symmetry allows nonzero Berry curvature (\textbf{$\Omega$}).
	Additionally, imaging of small-angle twisted \moire devices reveals strain that breaks three-fold rotational ($C_3$) symmetry~\cite{mcgilly_visualization_2020, huang_giant_2021, kazmierczak_strain_2021}.
	Such reduced symmetry makes the Berry curvature distribution non-symmetric and anisotropic over the \moire Brillouin zone (mBZ), leading to nonzero BCD ($\Lambda_{\alpha}$) given by,
	
	\begin{equation}\label{eqn:1}
		\Lambda_{\alpha} = \sum_{n} \int_{\rm mBZ}  \dfrac{d{\bf k}}{(2\pi)^2} \Omega_z^n \frac{\partial\epsilon^n_{\bf k}}{\hbar \partial k_{\alpha}} \frac{\partial f(\epsilon^n_{\bf k})}{\partial \epsilon^n_{\bf k}}.
	\end{equation}
	Here $\alpha$ stands for the spatial index $(x,y)$, $\epsilon_{\bf k}^n$ is the energy of the $n^{th}$ band and $f(\epsilon^n_{\bf k})$ is the Fermi-Dirac function. 
	In Eq.~\eqref{eqn:1}, a sum over all the bands crossing the Fermi energy is implied. 
	To calculate the BCD in TDBG, we consider distorted hexagonal \moire Brillouin zone with the $C_3$ symmetry broken by strain as shown in the schematic of Fig.~\ref{fig:fig1}a.
	The effect of $C_3$ symmetry breaking is reflected in the Berry curvature plots of Fig.~\ref{fig:fig1}b, c.
	Figure~\ref{fig:fig1}d shows the band structure for TDBG with a twist angle of 1.10$^\circ$ under 0.1\% strain with $\Omega_z$ of the flat bands indicated by the color.
	Together, nonzero Berry curvature and broken $C_3$ symmetry generate a nonzero BCD, as shown in Fig.~\ref{fig:fig1}e (see Methods and Supplementary Information section I-III for calculations).

	We fabricate multiple TDBG devices to measure NLH voltage, which probes the BCD~\cite{fu_quantum_2015,ma_observation_2019}. 
	The dual-gate geometry allows independent control of the charge density $(n)$ and the perpendicular electric displacement field $(D)$ (see Methods).
	Figure~\ref{fig:fig1}f shows our measurement schematic to probe NLH effect. 
	We send a current with low-frequency $\omega$ and measure the longitudinal voltage (\Vxxw) and the second-harmonic NLH voltage (\Vxytw) at frequencies $\omega$ and $2\omega$, respectively.
	In Fig.~\ref{fig:fig1}g, we show \Vxxw{} and \Vxytw{} as a function of \moire filling factor $(\nu)$ at zero displacement field for a TDBG device with twist angle 1.10$^\circ$. 
	Here $\nu=4 n/n_s$ with $n_s$ representing the number of carriers to fill one \moire band and factor 4 due to four-fold degeneracy of spin and valley.
	Peaks in \Vxxw{} at $\nu=\pm 4$ correspond to the two \moire gaps. 
	The substructures in the peaks of \Vxxw{} (indicated by arrows) result from slightly different twist angles, indicating an angle inhomogeneity of $\sim0.03^\circ$ -- a signature of strain~\cite{kazmierczak_strain_2021} that causes BCD in our system.
	In Fig.~\ref{fig:fig1}h, we plot \Vxxw{} and \Vxytw{} for a non-zero perpendicular displacement field that creates a gap at $\nu=0$ (charge neutrality point, CNP).
	We find NLH voltage \Vxytw{} is high in the vicinity of band edge around the \moire gaps at $\nu=\pm4$ and the CNP gap in our strained TDBG devices.

\begin{figure*}
		
		\centering
		\includegraphics[width=15.5cm]{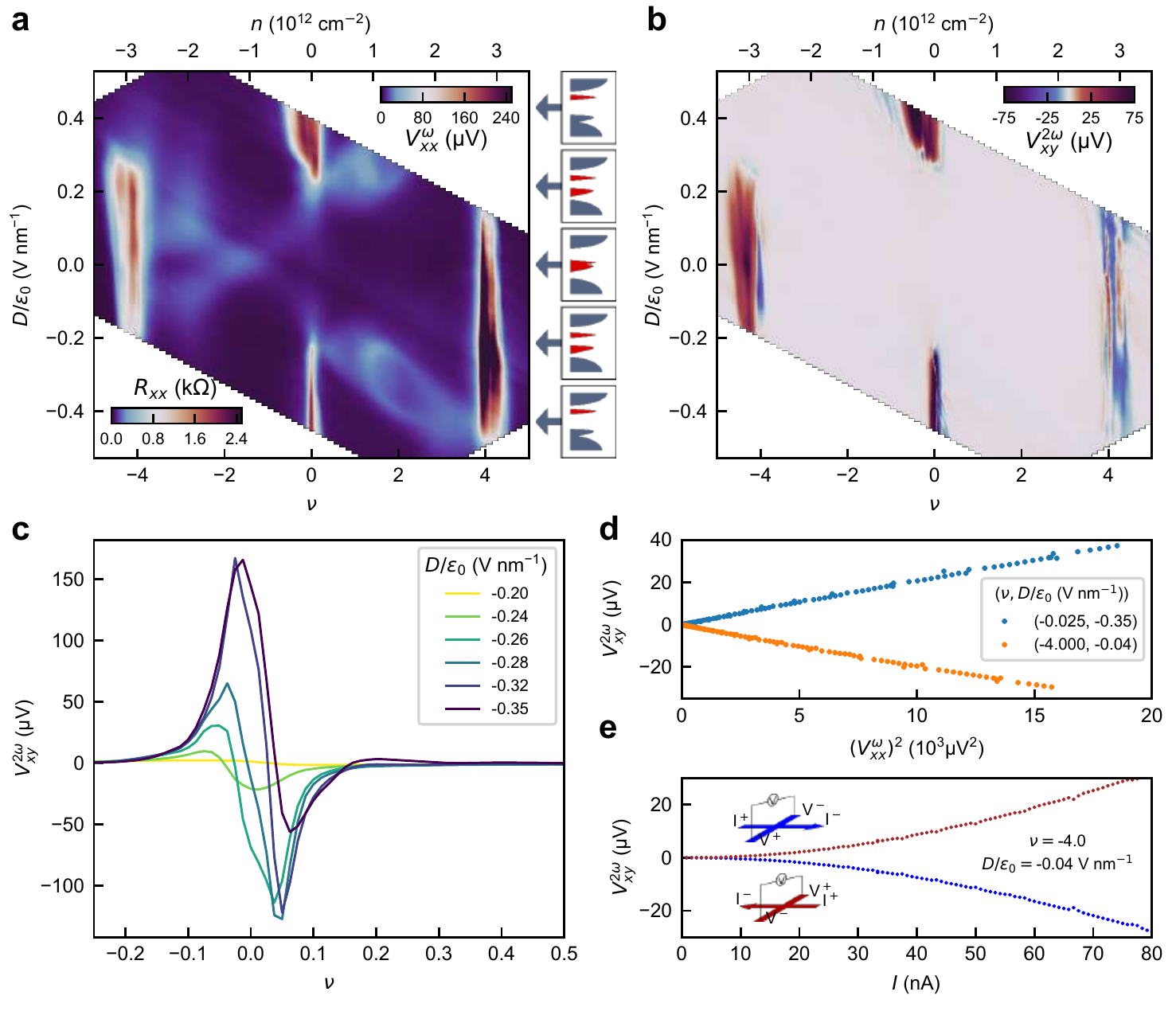}
		\caption{ \label{fig:fig2} { \textbf{Longitudinal and NLH voltages in TDBG.} \textbf{a,}~Longitudinal voltage (\Vxxw) as a function of filling factor ($\nu$) and perpendicular electric displacement field ($D$) measured by sending an ac current of $I= 100$~nA with frequency $\omega=177$~Hz at a temperature of 1.5~K. The top $x$-axis indicates the charge density ($n$). The color bar provided in bottom left of \textbf{a} indicates the corresponding values of the longitudinal resistance ($R_{xx}$). The schematics to the right of \textbf{a} depict the cartoon band structures with energy in the $y$-axis and the density of states (DOS) in the $x$-axis for different regimes of $D$ indicated by the arrows. The isolated flat bands are marked in red, while the dispersive \moire bands are marked in blue.
		\textbf{b,}~Second-harmonic NLH voltage (\Vxytw) as a function of $\nu$ and $D$ measured simultaneously with \Vxxw{} in \textbf{a}. \Vxytw{} peaks near the CNP gap and the \moire gaps.  
		\textbf{c,}~Line slices of \Vxytw{} vs. $\nu$ for different values of $D$. 
		\textbf{d,}~Linear dependence of \Vxytw{} on simultaneously measured (\Vxxw)$^2$ by varying current as a parameter upto 80~nA at $T=1.5$~K for two biasing points of ($\nu$,$D$). 
		\textbf{e,}~\Vxytw{} vs. $I$ for $\nu = -4$ and $D/\epsilon_0 = -0.04$~\Vbynm{}. The colored schematics indicate the orientation of the current and the voltage terminals. \Vxytw{} reverses sign when the orientation of both the voltage probes and the current probes are flipped simultaneously indicating the second-order nature of the NLH voltage. 
		}}   
	\end{figure*}

	We now study the evolution of \Vxxw{} and the corresponding resistance ($R_{xx}$) in the experimentally accessible parameter space of $\nu$ and $D$ in Fig.~\ref{fig:fig2}a. 
	Schematics adjacent to Fig.~\ref{fig:fig2}a show change in the underlying band structure with $D$ (see Supplementary Information section III for calculated band structure). 
	A finite displacement field opens up a gap at CNP between the conduction and the valence flat bands. 
	These flat bands are separated from the remote \moire bands by two \moire gaps, which close at larger values of $D$. 
	Apart from the peaks in \Vxxw{} corresponding to the CNP gap at $\nu=0$ and the \moire gaps at $\nu=\pm 4$, we observe other features like the cross towards the hole side ($-4<\nu<0$) and the halo around $D/\epsilon_0=\pm 0.3$~\Vbynm{} towards the electron side ($0<\nu<4$). 
	These characteristic features of TDBG within partial fillings of the flat bands are likely to be connected to structures in the density of states~\cite{he_symmetry_2020}. 
	
\begin{figure*}
		\centering
		\includegraphics[width=16cm]{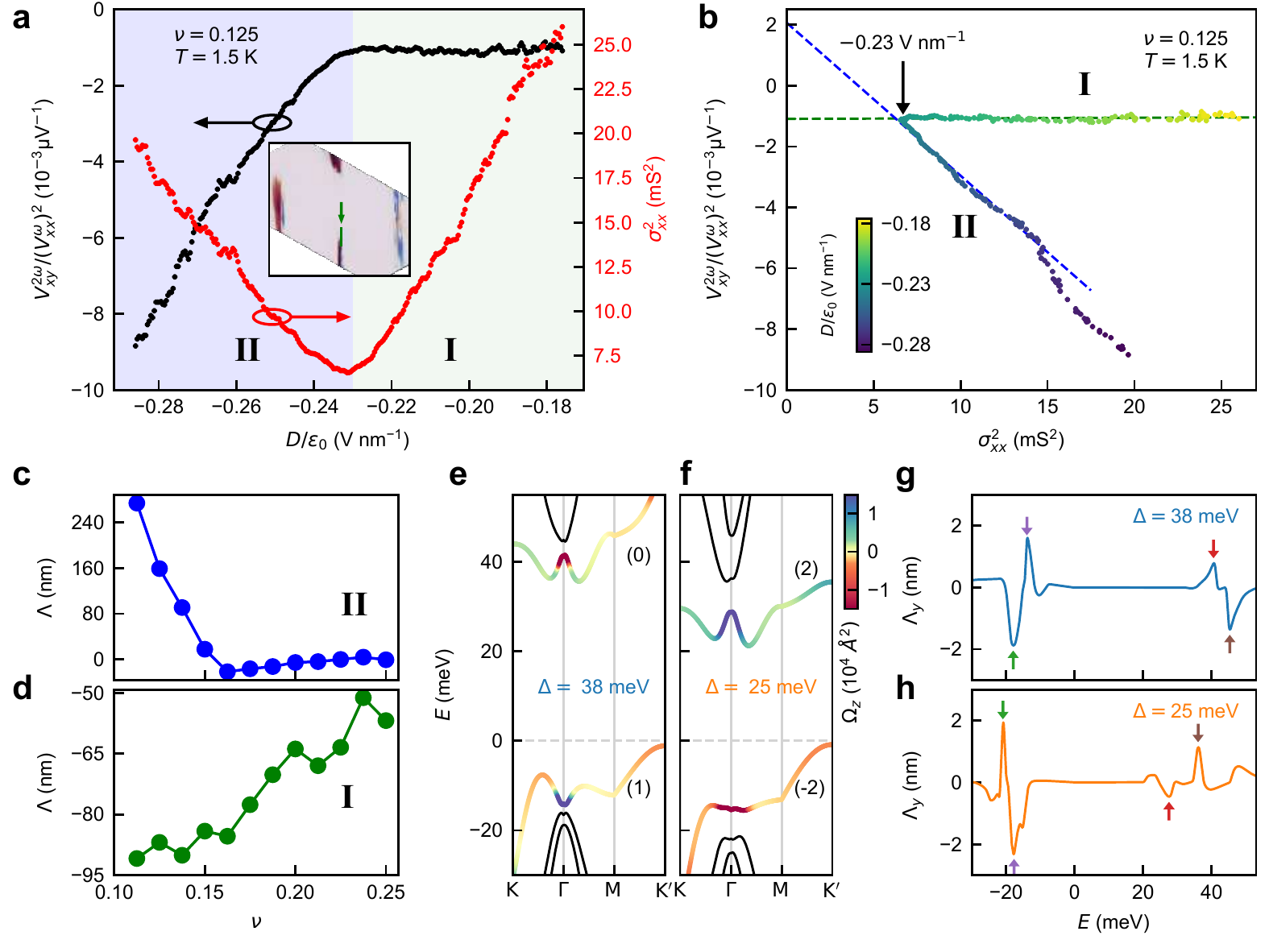}
		\caption{ \label{fig:fig3} { \textbf{Estimation of Berry curvature dipole (BCD) and its sign change across topological transition.} 
				\textbf{a,}~Dependence of normalized NLH voltage \Vxytw/(\Vxxw)$^2$ (black data points, left axis) and square of longitudinal conductivity $\sigma_{xx}^2$ (red data points, right axis) on the displacement field ($D$). $D$ is slowly varied at 1.5~K with the filling factor fixed at $\nu=0.125$, which places the Fermi energy in the flat conduction band. 
				The green vertical line pointed by the arrow in the inset of \Vxytw{} vs. $\nu$ and $D$ (as in Fig.~\ref{fig:fig2}b) marks the $D$-range used for the plot. 
				The valley Chern numbers of the bands change across the two regimes of $D$ (regime-I with light-green background and regime-II with light-blue background), separated at $D/\epsilon_0=-0.23$~V~nm$^{-1}$.
				\textbf{b,}~Scaling of \Vxytw/(\Vxxw)$^2$ with $\sigma_{xx}^2$ using $D$ as a parameter for the same $\nu$ and $T$ as in \textbf{a}.
				The color of the data points indicates the corresponding $D/\epsilon_0$ values, where $D$ is swept at a slow rate. The vertical arrow marks the transition point between regime-I ($|D|/\epsilon_0$<0.23~V~nm$^{-1}$) and regime-II ($|D|/\epsilon_0$>0.23 V~nm$^{-1}$). The $y$-intercept of the linear fit gives the estimate of the BCD.
				\textbf{c, d,}~Extracted BCD ($\Lambda$) at different values of $\nu$  close to the CNP ($\nu=0$) for regime-II (\textbf{c}) and regime-I (\textbf{d}). 
				The fitting error bars are smaller than the size of the individual data points. 
				\textbf{e, f,}~The band structures of $K$ valley for two different values of the inter-layer potential $\Delta=38$~meV (\textbf{e}) and $\Delta=25$~meV (\textbf{f}). The overlayed color indicates the Berry curvature for the corresponding flat bands. The Berry curvature, and consequently, the valley Chern numbers for the bands (indicated within brackets) change across the topological transition. 
				\textbf{g, h,}~The corresponding variation of the calculated BCD as a function of energy ($E$) for $\Delta=38$~meV (\textbf{g}) and $\Delta=25$~meV (\textbf{h}). For a fixed band filling, the BCD changes sign across the topological transition. The green, purple, red, and brown arrows point to the sequence of the BCD peaks with the increase in $E$ values.}}
\end{figure*}

	Figure~\ref{fig:fig2}b shows the evolution of the NLH voltage \Vxytw{} in the same parameter space of $\nu$ and $D$.
	In Fig.~\ref{fig:fig2}c, we have plotted few line slices of \Vxytw{} around $\nu=0$ for different $D$.
	We find that \Vxytw{} peaks near the gaps at $\nu=0$ (CNP gap) and $\nu=\pm 4$ (\moire gaps).
	The coincidence of the NLH voltage with the TDBG bandgaps reveals that the Berry curvature hotspots, and hence the BCD, reside predominantly in the vicinity of the band edges.
	We explore this behavior further in Fig.~\ref{fig:fig3}.
	
	\Vxytw{} that we measure should scale quadratically with the in-plane ac electric field of frequency $\omega$. 
	To verify the quadratic scaling, we plot \Vxytw{} against  the square of simultaneously measured \Vxxw{} for two different ($\nu, D$) biasing points in Fig.~\ref{fig:fig2}d. 
	The linear behavior of \Vxytw{} with the square of \Vxxw{} confirms the quadratic scaling near both the CNP gap and the \moire gaps.   
	Additionally, \Vxytw{} switches sign when we reverse the Hall voltage probes and the direction of current simultaneously as depicted in Fig.~\ref{fig:fig2}e. 
	These observations clearly demonstrate the second-order nature of the measured NLH voltage, \Vxytw{}.

	Having established \Vxytw{} as NLH voltage, we now use $D$ as a parameter to study scaling between normalized NLH voltage \Vxytw/(\Vxxw)$^2$ and square of longitudinal conductivity ($\sigma_{xx}^2$).
	Such scaling can be used to quantify BCD, as discussed later.
	In Fig.~\ref{fig:fig3}a, we plot \Vxytw/(\Vxxw)$^2$ (left axis) and $\sigma_{xx}^2$ (right axis) as a function of $D$ for $\nu=0.125$ (indicated by the green vertical line in inset of Fig.~\ref{fig:fig3}a). 
	The metallic behavior at $\nu=0.125$ ensures that the Fermi energy is near the band edge (see Extended Data Fig.~3a for the temperature ($T$) dependence of $R_{xx}$).
	Both \Vxytw/(\Vxxw)$^2$ and $\sigma_{xx}^2$ show different trends in two different regimes of the displacement field separated at $D/\epsilon_0=-0.23$~\Vbynm{}. 
	Interestingly, the two regimes are characterized by different band structures-- in regime-I, the flat bands are gapped and are isolated from the remote \moire bands, while in regime-II the valence flat band merges with the remote \moire band (see Methods and Extended Data Fig.~3b, c). 
	It is known that changes in band structure due to $D$ lead to change in the valley Chern numbers of the bands~\cite{koshino_band_2019-1,zhang_nearly_2019} which in turn should be reflected in the change in sign of the BCD~\cite{hu_nonlinear_2020}.
	
	To investigate the close connection of such topological transition with BCD, we first experimentally analyze  the scaling between \Vxytw/(\Vxxw)$^2$ and $\sigma_{xx}^2$ and later present a detailed calculation.  
	In Fig.~\ref{fig:fig3}b, we find linear scaling between \Vxytw/(\Vxxw)$^2$ and $\sigma_{xx}^2$ in both regime-I and regime-II.
	The slope of a linear \Vxytw/(\Vxxw)$^2$ vs. $\sigma_{xx}^2$ dependence characterizes the contribution from skew-scattering ($\tau^3$) process~\cite{he_quantum_2021}, while the BCD can be extracted from the $y$-intercept~\cite{kang_nonlinear_2019,he_quantum_2021} (see Supplementary Information section VI).
	A fixed temperature assures that the contributions from skew-scattering or side-jump in the NLH effect are not tuned within the linear fitted regimes~\cite{xiao_scaling_2019,du_disorder-induced_2019}.
	Interestingly, near-zero slope in regime-I suggests that the skew-scattering contribution is minimal when the flat bands are isolated from the remote \moire bands.
	
	To distinguish the intrinsic BCD from other extrinsic contributions like side-jump, we repeat the scaling analysis for different fillings close to the band edge (see Extended Data Fig.~1).
	Figures~\ref{fig:fig3}c and 3d show the extracted BCD from regime-II and regime-I respectively.
	The drop in extracted BCD away from CNP ($\nu=0$), consistent with Berry curvature hotspot peaking at the band edge, suggests that the NLH effect we observe is BCD dominated.
	The most striking observation, central to our study, is the sign change of BCD across the topological transition between regime-I and regime-II.
	
	To theoretically verify the connection between BCD sign change and the topological transition, we calculate the band structure for two different values of inter-layer potential in Fig.~\ref{fig:fig3}e, f. 
	A change in valley Chern number marks the topological transition.
	Consistently, the BCD changes sign as seen in Fig.~\ref{fig:fig3}h and \ref{fig:fig3}g before and after the topological transition, respectively (also see Extended Data Fig.~4).
	We also note from Fig.~\ref{fig:fig3}e and \ref{fig:fig3}f that the Berry curvature changes sign, around the $\Gamma$-point, across the topological transition.
	Thus, the NLH effect acts as a good probe to detect topological transitions associated with the change in the valley Chern numbers~\cite{facio_strongly_2018}.

	We note that the BCD values in Fig.~\ref{fig:fig3}c, d are higher than the BCD observed in WTe$_2$~\cite{kang_nonlinear_2019,ma_observation_2019}. 
	The scaling analysis using temperature as a parameter also gives a similar magnitude of BCD (see Extended Data Fig.~2). High BCD magnitude has been theoretically predicted in strained twisted systems~\cite{hu_nonlinear_2020,huang_giant_2021, pantaleon_tunable_2021,zhang_giant_2020, he_giant_2021}. 
	The origin of high BCD in graphene-based \moire systems, in the vicinity of the charge neutrality point, can be understood using the model of strained twisted bilayer graphene as a tilted Dirac system~\cite{mannai_twistronics_2021}.
	In this model, BCD $\propto t \propto \lambda^3$, where $t$ is the tilt parameter and $\lambda$ is the \moire wavelength.
	So, the BCD is expected to be higher in \moire materials as compared to non-\moire materials like bilayer graphene, etc. (see Supplementary Information section V for details).

	\begin{figure*}
		\centering
		\includegraphics[width=15cm]{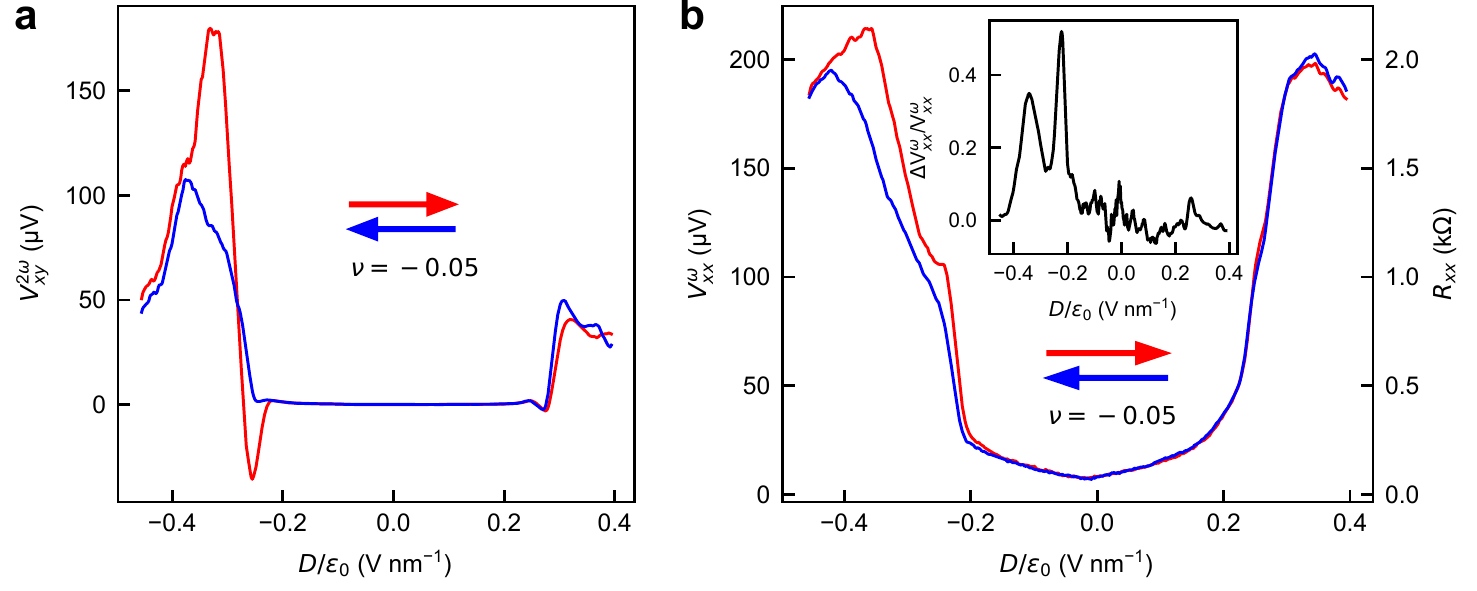}
		\caption{ \label{fig:fig4} { \textbf{Hysteresis in the NLH and longitudinal voltages.} 
				\textbf{a,}~Variation of the NLH voltage \Vxytw{} with $D$ for a filling $\nu=-0.05$ around CNP at 1.5~K. The arrows indicate the sweep direction of $D/\epsilon_0$. 
				\textbf{b,}~Hysteresis in \Vxxw{} for up and down sweeps of $D$ taken simultaneously with \textbf{a}. The right axis indicates the corresponding longitudinal resistance, $R_{xx}$. The inset shows the relative difference in \Vxxw{} of up and down sweep, obtained by averaging over three consecutive data points. The relative difference is maximum around $D/\epsilon_0=-0.23$~\Vbynm{}. The displacement field was swept at a constant rate of 2.5~mV nm$^{-1}$ s$^{-1}$.}}
	\end{figure*}
	Finally, we examine the variation of NLH voltage \Vxytw{} at large displacement fields and, interestingly, observe a hysteretic response as shown in Fig.~\ref{fig:fig4}a.
	Fig.~\ref{fig:fig4}b shows the simultaneously measured \Vxxw{}, which also reveals hysteresis with $D$.
	The sense of hysteresis is doping dependent (see Supplementary Information section X.4) and cannot be explained by charge traps in dielectric.
	See Supplementary Information section X for temperature effects, rate dependence, and switching statistics of the hysteresis.
	The inset of Fig.~\ref{fig:fig4}b shows the difference in \Vxxw{} for up and down sweeps.
	Intriguingly, the hysteresis is maximal around $D/\epsilon_0=-0.23$~\Vbynm{} where a topological transition takes place as shown in Fig.~\ref{fig:fig3}b.
	However, more studies are required for a full microscopic understanding.

	Our experimental data suggest the existence of metastable states; the origin of these states can be due to two possible mechanisms.
	Firstly, metastable polarization states in the system can arise from a difference in charge density across layers due to symmetry breaking in the device geometry (see Methods for further discussion).  
	Secondly, flexoelectric coupling between perpendicular electric field and the strain gradient across the domain interfaces can lead to electric polarization in \moire systems~\cite{li_unraveling_2021}. 
	Polarization from both possible mechanisms can lead to change in strain profile that can be further pinned by defects; this can account for the hysteresis we observe. 
	Strain and polarization, through the effective electric field, modify the band structure and hence the value of \Vxxw{} and \Vxytw. 
	
	Tunable BCD values, together with the coupling of BCD to hysteresis in the same platform, as we demonstrate using TDBG, may open new frontiers for next-generation memory devices.
	In general, probing topological transitions using electrical transport is challenging.
	Our experiment demonstrates that the NLH effect can probe topological transitions.
	Electric field tunable band structure of few-layer heterostructures provides a useful way for probing topological transitions, as we show using TDBG. 
	Our approach to detect topological transitions using NLH effect is applicable for 3D systems as well.

	\section*{Acknowledgements:}
	We thank Justin Song, Abhay Pasupathy, Senthil Todadri, Biswajit Datta, Sanat Ghosh, and Supriya Mandal
	for helpful discussions and comments. We thank Surya Kanthi R.~S., Joydip Sarkar, Krishnendu Maji, and Raghav Dhingra for experimental assistance. M.M.D. acknowledges Nanomission grant SR/NM/NS-45/2016 and DST SUPRA SPR/2019/001247 grant along with Department of Atomic Energy of Government of India 12-R\&D-TFR-5.10-0100 for support.
	K.W. and T.T. acknowledge support from the Elemental Strategy Initiative conducted by the MEXT, Japan (Grant Number JPMXP0112101001) and  JSPS KAKENHI (Grant Numbers 19H05790 and JP20H00354).
	A.A. acknowledges IIT Kanpur - India, Science Engineering and Research Board (SERB)- India, and the Department of Science and Technology (DST) - India, for financial support. 
	A.C. acknowledges the Institute Post-Doctoral fellowship of IIT Kanpur.
	K. Das acknowledges IIT Kanpur for Senior Research Fellowship.
	A.A., A.C., and K. Das also thank CC-IIT  Kanpur, for the high-performance computing facility.
	K. Debnath is grateful to the Jawaharlal Nehru Centre for Advanced Scientific Research, India, for a research fellowship. 
	U.V.W. acknowledges support from a JC Bose National Fellowship of SERB-DST.

	\section*{Data Availability:}
	The experimental data used in the figures of the main text are available in Zenodo with the identifier \href{https://doi.org/10.5281/zenodo.5211285}{doi:10.5281/zenodo.5211285}~\cite{sinha_berry_2021}.
	Additional data related to this study are available from the corresponding authors upon reasonable request.

	\section*{Author Contributions:}
	S.S., P.C.A., and L.D.V.S. fabricated the devices. S.S. and P.C.A. did the measurements and analyzed the data. A.C., K. Das, and A.A. did the band structure, BCD and Chern number calculations. K. Debnath and U.V.W. did the polarization calculations. K.W. and T.T. grew the hBN crystals. S.S., P.C.A., A.A., and M.M.D. wrote the manuscript with inputs from everyone. M.M.D. supervised the project.
	
	\section*{Competing Interests:}
	The authors declare no competing interests.
	

\clearpage
\newpage
\onecolumn
\begin{center}
	\large \textbf{METHODS}
\end{center}

\setcounter{figure}{0}

\captionsetup[figure]{labelfont={bf},labelsep=period,name={Extended Data Fig.}}

\section{Device fabrication}
We fabricate multiple dual gated TDBG devices. 
We have a metal top gate and SiO$_2$($\sim$280~nm)/Si$^{++}$ back gate. 
Dual gates allow us to independently control the charge density ($n$) and the perpendicular displacement field ($D$), where $n=(C_{BG} V_{BG}+C_{TG} V_{TG})/e$ and $D=(C_{BG} V_{BG}-C_{TG} V_{TG})/2$. 
Here, $C_{BG} (C_{TG})$ is the back (top) gate capacitance and $V_{BG} (V_{TG})$ is the  voltage applied to the back (top) gate, respectively. $e$ is the electronic charge.
To prepare the hBN/TDBG/hBN graphene stacks, we pre-cut bilayer graphene using a tapered optical fiber scalpel~\cite{varma_sangani_facile_2020}.
The detail fabrication and twist angle determination procedure is discussed in the methods of Ref.~\cite{sinha_bulk_2020-4}.
The twist angle is additionally confirmed via magneto-transport measurements.

\section{Measurement technique}
We send current using DS360 ultra-low distortion function generator.
The chosen frequency for all the data is 177~Hz, except for Supplementary Fig.~8 where we vary the frequency.
The voltages, \Vxxw{} and \Vxytw, are measured using SR830 lock-in amplifier in the first harmonic and second harmonic mode, respectively.
In addition, to measure the DC voltage (Supplementary Fig.~20) generated due to the NLH effect, we use Keithley 2182 nanovoltmeter.
For applying DC voltages to the top and back gate, we use NI DAQ.
The temperature is measured using a sensor placed close to the device.
The findings reported in the main manuscript are from a TDBG device with twist angle 1.10$^\circ$; other devices are shown in Supplementary Fig.~5 and 16.

\section{BCD estimation}
\subsection{Using electric field as parameter}
To extract out the BCD ($\Lambda$), we use the formula $\frac{E_\perp^{2\omega}}{(E_\parallel)^2}=\xi \sigma_{xx}^2+\eta$, where $E_\perp^{2\omega}$ is the second-order nonlinear electric field, $E_\parallel$ is the in-plane longitudinal electric field, parallel to the direction of current, $\xi$ is the slope and $\eta$ is the intercept.
Using the fact that the length and width of our device are 2~\textmu m each, $\eta=2 \cross$ $y$-intercept of $\frac{V_{xy}^{2\omega}}{(V_{xx}^{\omega})^2}$ vs. $\sigma_{xx}^2$ dependence.
We fit the $D$-parametric plot of $\frac{V_{xy}^{2\omega}}{(V_{xx}^{\omega})^2}$ vs. $\sigma_{xx}^2$ with a linear dependence (see Supplementary Information section VI for details) to extract $\eta$ and use the relation $\Lambda=\eta \frac{E_F}{\pi e}$ to estimate the BCD~\cite{kang_nonlinear_2019}.
We assume that $E_F$ is $\sim$1\% (limited by the energy scale $k_BT$ set by the temperature $T=$~1.5~K) of the bandwidth ($\sim$10~meV) of the flat band. 
To ensure that we place the Fermi energy $E_F$ within the flat band and not at the gap, we measure the temperature dependence of $R_{xx}$ and find a metallic behavior irrespective of the chosen value of $D$ (see Extended Data Fig.~\ref{fig:RvsT}a).

\begin{figure*}
	\centering
	\includegraphics[width=16cm]{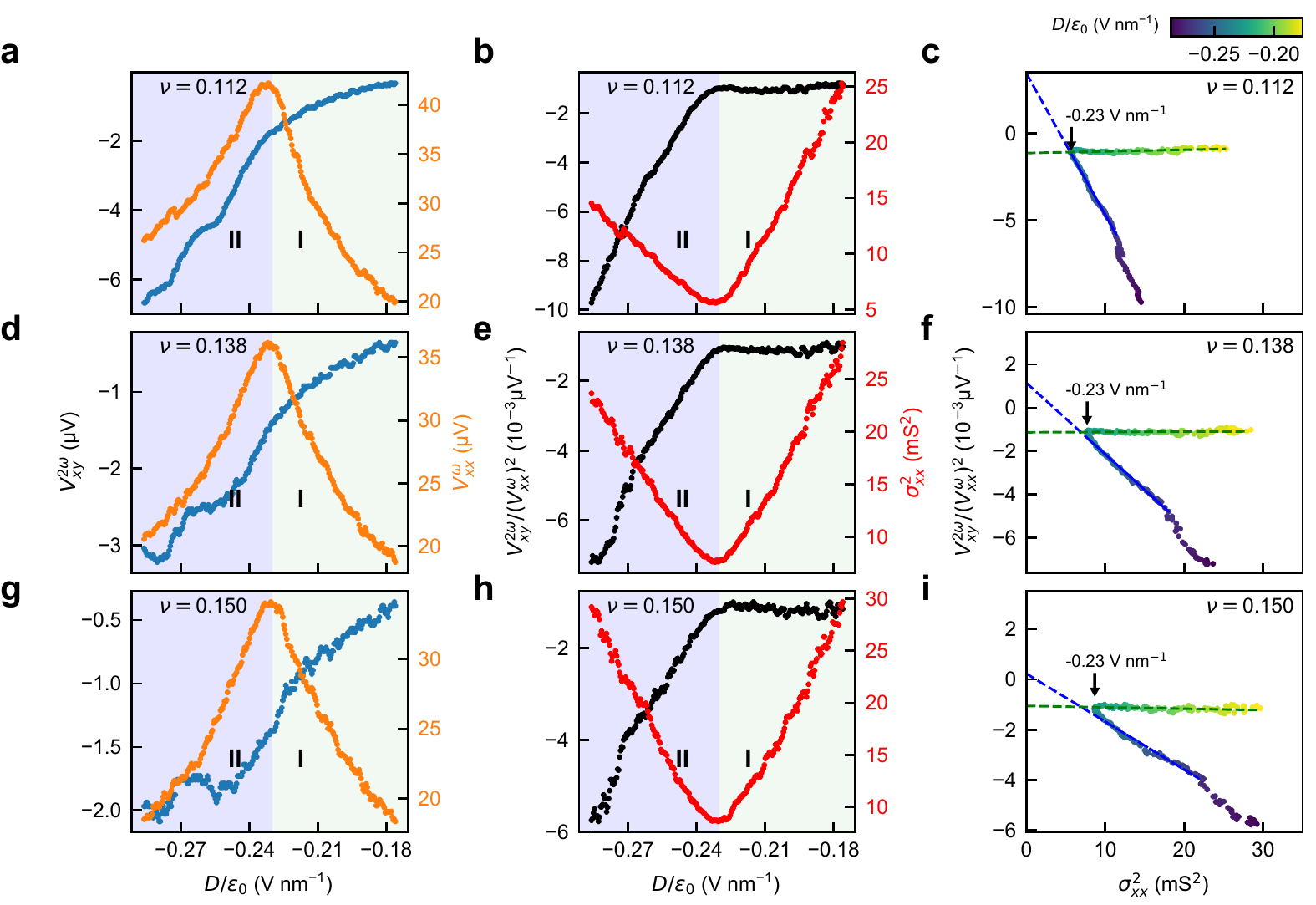}
	\caption{ \label{fig:Dparameter} {\textbf{Scaling of normalized nonlinear Hall voltage \Vxytw/(\Vxxw)$^2$ with the square of longitudinal conductivity ($\sigma_{xx}^2$) for different fillings $\nu$ with displacement field as parameter.}}
		\textbf{a, d, g,}~The variation of nonlinear Hall voltage \Vxytw{} (blue-colored data points corresponding to the left axis) and longitudinal voltage \Vxxw{} (orange-colored data points corresponding to the right axis) as a function of the displacement field $D/\epsilon_0$ for three different fillings.
		\textbf{b, e, h,}~The corresponding variation of normalized nonlinear Hall voltage \Vxytw/(\Vxxw)$^2$ (black colored data points corresponding to the left axis) and square of longitudinal conductivity $\sigma_{xx}^2$ (red-colored data points corresponding to the right axis) as a function of the displacement field $D/\epsilon_0$, extracted for the same fillings used in \textbf{a}, \textbf{g} and \textbf{d}, respectively.
		\textbf{c, f, i,}~The variation of normalized nonlinear Hall voltage \Vxytw/(\Vxxw)$^2$ with square of longitudinal conductivity $\sigma_{xx}^2$ plotted parametrically as a function of the displacement field $D/\epsilon_0$, using \textbf{b}, \textbf{e} and \textbf{h}, respectively.
		The displacement field value of data points in \Vbynm{} is indicated by the color (color bar is shown in top right).
		The dashed green line and dashed blue line indicate fits to linear scaling in regime-I and regime-II respectively, used to extract BCD. 
		The fillings shown here are $\nu=$~0.112 (\textbf{a-c}), 0.138 (\textbf{d-f}) and 0.150 (\textbf{g-i}). The light green background and light blue background correspond to regime-I and regime-II, respectively, as discussed in Fig.~3a of the main manuscript.
		The measurements were performed using a current of 100~nA with a frequency of 177~Hz at a temperature of 1.5~K. 
	}
\end{figure*}

To verify the robustness of our observation of BCD sign change, we systematically show the scaling of normalized nonlinear Hall voltage (\Vxytw/(\Vxxw)$^2$) with square of longitudinal conductivity ($\sigma_{xx}^2$) for three different fillings ($\nu$) in Extended Data Fig.~\ref{fig:Dparameter}.
Extended Data Fig.~\ref{fig:Dparameter}a, d, and g shows the measured nonlinear Hall voltage \Vxytw{} and longitudinal voltage \Vxxw{} as a function of the displacement field for three different $\nu$. 
The background color denotes the two regimes-- regime-I and regime-II, across which a topological transition occurs.
We see broad peaks in \Vxxw{} across the critical electric field of $D/\epsilon_0=-0.23$~\Vbynm{} in all three cases.
Using the measured \Vxxw{}  and \Vxytw{}  as in Extended Data Fig.~\ref{fig:Dparameter}a, d, and g, we plot the corresponding normalized NLH voltage \Vxytw/(\Vxxw)$^2$ and $\sigma_{xx}^2=$~(I/\Vxxw)$^2$ as a function of displacement field for the three fillings in Extended Data Fig.~\ref{fig:Dparameter}b, e, and h.
In Extended Data Fig.~\ref{fig:Dparameter}c, f, and i, we plot \Vxytw/(\Vxxw)$^2$ vs. $\sigma_{xx}^2$, using displacement field as a parameter.
The displacement field ($D/\epsilon_0=-0.23$~\Vbynm), at which the change in sign of intercept occurs, remains invariant for different fillings. 
We extract the intercept and use it to estimate BCD as described above, and plot in Fig.~3c and 3d of the main manuscript.

The extracted BCD values from the two regimes are opposite in sign.
A characteristic signature of a topological transition is a change in sign of BCD, as discussed in the main manuscript and also in Ref.~\cite{hu_nonlinear_2020} and Ref.~\cite{facio_strongly_2018}.
As we show via resistance vs. temperature measurements in Extended Data Fig.~\ref{fig:RvsT}c, displacement field closes the hole-side \moire gap around the same $D/\epsilon_0=-0.23$~\Vbynm.
Displacement field, that causes the flat bands to touch the remote \moire valence bands, changes the valley Chern number of the bands in TDBG~\cite{koshino_band_2019-1, zhang_nearly_2019}, as we independently show in our calculations (Fig.~3e,f of the main manuscript) for the twist angle of 1.1$^\circ$.
Such a change in sign of valley Chern number accounts for the two linear regimes with different signs of intercepts.

\subsection{Using temperature as parameter}
\begin{figure*}
	\centering
	\includegraphics[width=16cm]{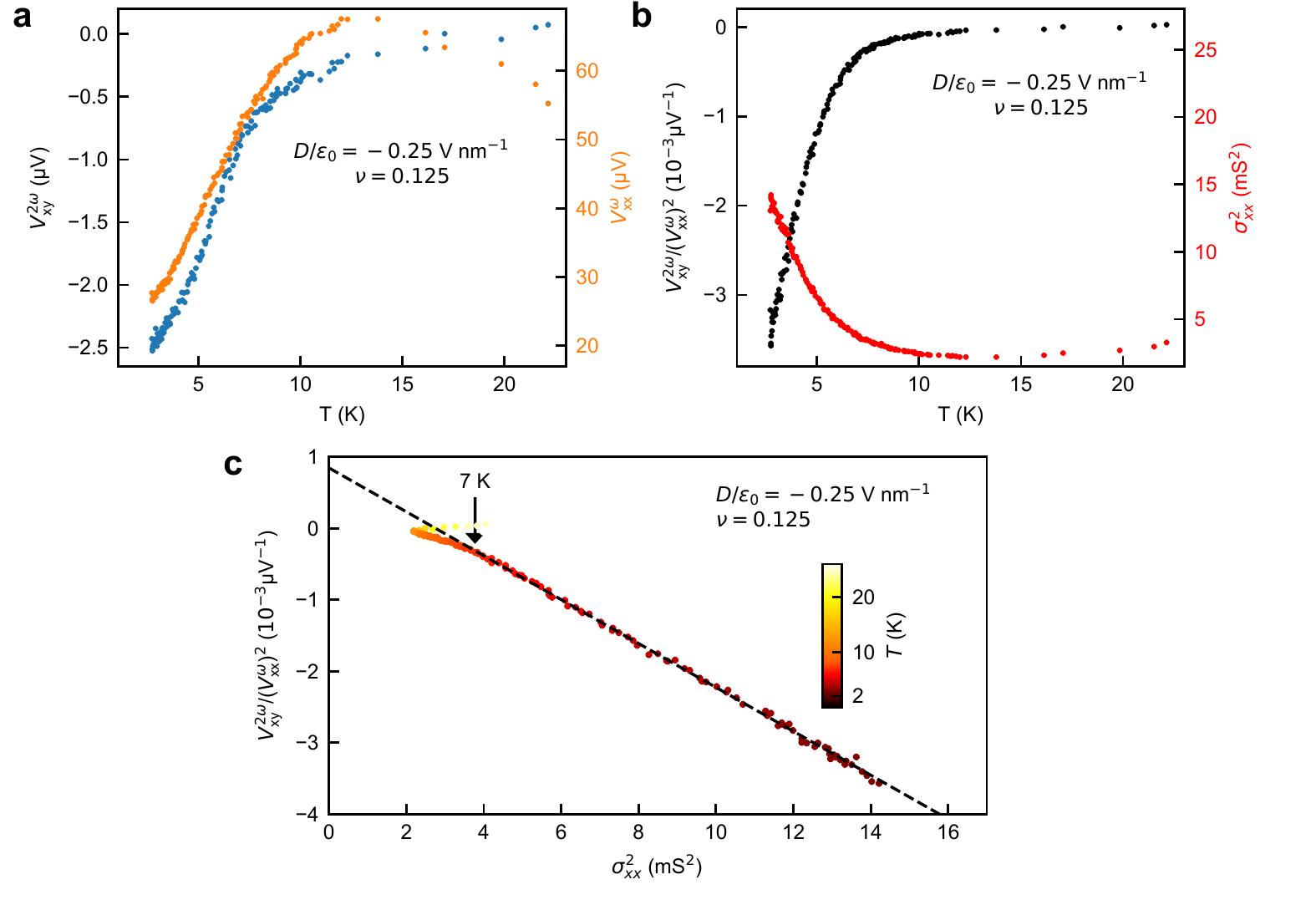}
	\caption{ \label{fig:Temppara} { \textbf{Scaling of normalized nonlinear Hall voltage \Vxytw/(\Vxxw)$^2$ with square of longitudinal conductivity ($\sigma_{xx}^2$) with temperature as parameter.}}
		\textbf{a,}~The variation of nonlinear Hall voltage \Vxytw{} (blue-colored data points corresponding to the left axis) and longitudinal voltage \Vxxw{} (orange-colored data points corresponding to the right axis) as a function of temperature $T$ for $\nu=0.125$.
		\textbf{b,}~The corresponding variation of normalized nonlinear Hall voltage \Vxytw/(\Vxxw)$^2$ (black-colored data points corresponding to the left axis) and square of longitudinal conductivity $\sigma_{xx}^2$ (red-colored data points corresponding to the right axis) as a function of $T$, extracted for the same filling used in \textbf{a}.
		\textbf{c,}~The variation of \Vxytw/(\Vxxw)$^2$ with $\sigma_{xx}^2$ plotted parametrically as a function of $T$, using the results in \textbf{b}.
		The temperature value of data points in Kelvin is indicated by the color.
		The dashed line indicates a linear fit till 7~K.
	}
\end{figure*}

To independently confirm the order of magnitude of the estimated BCD, we also study the temperature dependence of the NLH voltage.
In Extended Data Fig.~\ref{fig:Temppara}, we show the scaling of normalized nonlinear Hall voltage (\Vxytw/(\Vxxw)$^2$) with the square of longitudinal conductivity ($\sigma_{xx}$) for the filling $\nu=0.125$ (the same $\nu$ for which we show a linear scaling, using displacement field as a parameter, in Fig.~3b of the main manuscript) and displacement field $D/\epsilon_0=-0.25$~\Vbynm.
In Extended Data Fig.~\ref{fig:Temppara}a, we show the temperature dependence of the NLH voltage \Vxytw{} (blue-colored data points) and longitudinal voltage \Vxxw{} (orange-colored data points). 
We see that \Vxytw{} goes close to zero at $\sim$~20~K.
In Extended Data Fig.~\ref{fig:Temppara}b, we plot the temperature dependence of \Vxytw/(\Vxxw)$^2$ (black data points) and $\sigma_{xx}^2$ (red data points), using the data in Extended Data Fig.~\ref{fig:Temppara}a.
A decreasing $\sigma_{xx}^2$ with increasing temperature suggests that the filling $\nu=0.125$ corresponds to placing the Fermi level in the conduction flat band.
In Extended Data Fig.~\ref{fig:Temppara}c, we plot \Vxytw/(\Vxxw)$^2$ vs. $\sigma_{xx}^2$, using temperature as a parameter and find similar linear dependence, as reported earlier~\cite{kang_nonlinear_2019}, till $T=7$~K.
The temperature regime is less than the Bloch-Gr$\ddot{u}$neisen temperature, ensuring that the other extrinsic contributions are not tuned in this temperature regime.
A similar magnitude of $y$-intercept in \Vxytw/(\Vxxw)$^2$ as in Fig.~3b of the main manuscript, suggests that BCD using temperature as a parameter is of similar magnitude that we get from parametric scaling of $D$.

\section{Inferring band structure using R vs. T dependence}

\begin{figure*}
	\centering
	\includegraphics[width=16cm]{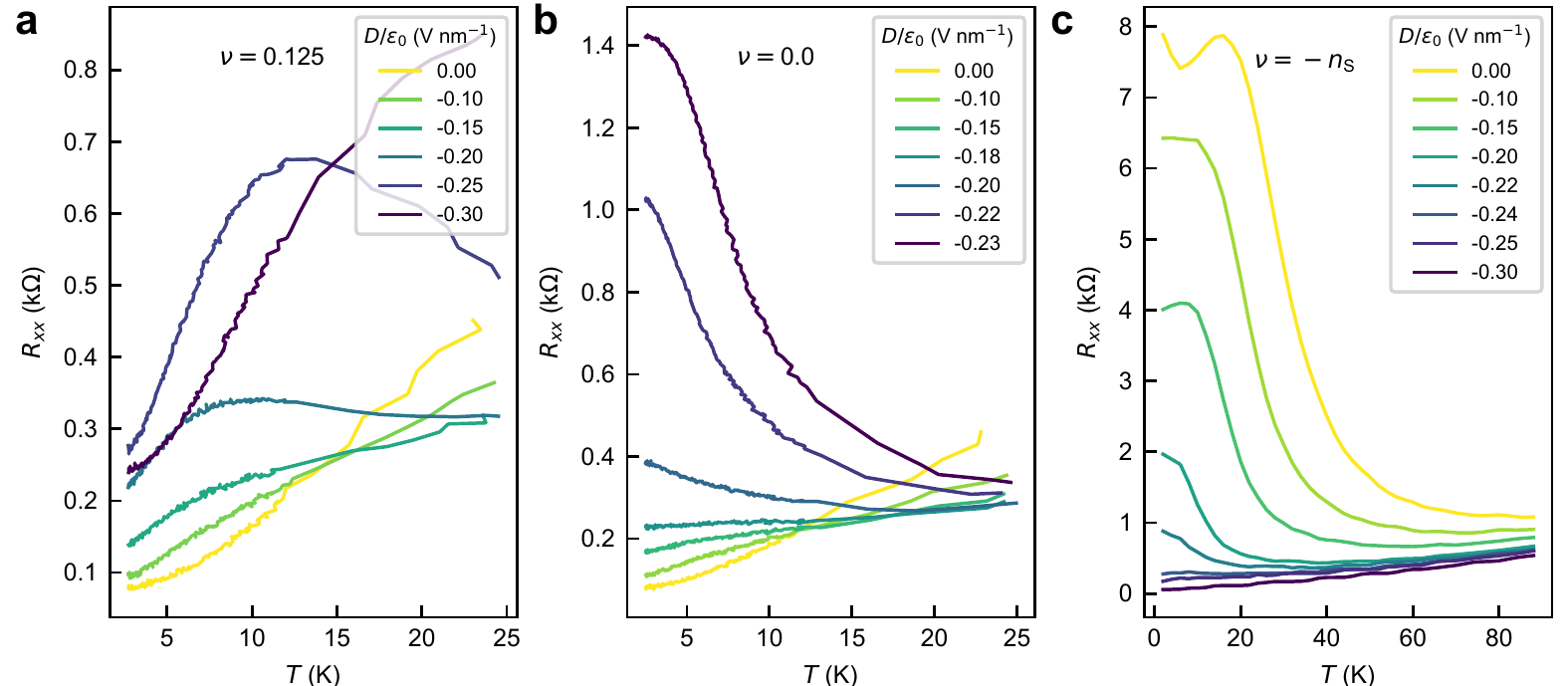}
	\caption{ \label{fig:RvsT} {\textbf{Variation of longitudinal resistance with temperature.}}
		\textbf{a-c,}~Variation of longitudinal resistance $R_{xx}$ with temperature $T$ for filling $\nu=0.125$ (\textbf{a}), $\nu=0$ (\textbf{b}) and $\nu=-4$ (\textbf{c}). The color of the line-plots indicates the corresponding displacement field. }
\end{figure*}

In Extended Data Fig.~\ref{fig:RvsT}, we show the temperature ($T$) variation of longitudinal resistance ($R_{xx}$) for fillings at $\nu=0$ (corresponding to CNP in Extended Data Fig.~\ref{fig:RvsT}b), $\nu=0.125$ (corresponding to conduction flat band edge in Extended Data Fig.~\ref{fig:RvsT}a) and $\nu=-4$ (corresponding to hole-side \moire gap in Extended Data Fig.~\ref{fig:RvsT}c) at different displacement fields ($D$).
In Extended Data Fig.~\ref{fig:RvsT}b, we see that as the magnitude of displacement field is increased, the slope of $R_{xx}$ vs. $T$ changes from positive to negative, indicating a gap opening at CNP at a displacement field around $|D|/\epsilon_0$ of 0.2~\Vbynm.
This establishes that in regime-I, a gap at CNP persists between the red colored flat bands.

In Extended Data Fig.~\ref{fig:RvsT}c, we see that as the magnitude of displacement field is increased, the slope of $R_{xx}$ vs. $T$ changes from negative to positive, indicating a gap closing at the hole-side \moire gap at $\nu=-4$ at a displacement field around $|D|/\epsilon_0$ of 0.23~\Vbynm.
This establishes that in regime-II, the hole-side \moire gap is closed and the flat band has merged with the hole-side remote \moire band.

\section{Berry curvature dipole and topological transition}
To investigate the topological phase-transition and associated sign reversal of Berry curvature dipole we have followed the low-energy continuum model approach of Bistritzer and MacDonald \cite{bistritzer_moire_2011} for twisted bilayer graphene and extended it to TDBG (see Supplementary Information section~I for details). The perpendicular electric field tunability of electronic band structure has been included in the Hamiltonian by a parameter $\Delta$, which represents constant potential gradient across the layers. The presence of finite strain in the fabricated TDBG sample breaks the $C_3$ symmetry (see Fig.~1a) which causes non-zero BCD.  The uniaxial strain has been introduced into our model Hamiltonian as~\cite{He2020}
\be \label{strain}
{\mathcal E} = \epsilon
\begin{pmatrix}
	-\cos^2 \phi + \nu \sin^2 \phi & -(1+ \nu) \sin \phi \cos \phi \\
	-(1+ \nu) \sin \phi \cos \phi & -\sin^2 \phi + \nu \cos^2 \phi  
\end{pmatrix},
\ee
where a strain of magnitude $\epsilon$ is applied along a direction making angle $\phi$ with the zigzag direction of graphene. $\nu$, the Poisson's ratio is equal to $\sim 0.16$ for graphene. For our calculations, we have considered strain along the zigzag direction ($\phi= 0^\circ$). 

\par
\begin{figure*}
	\centering
	\includegraphics[width=16cm]{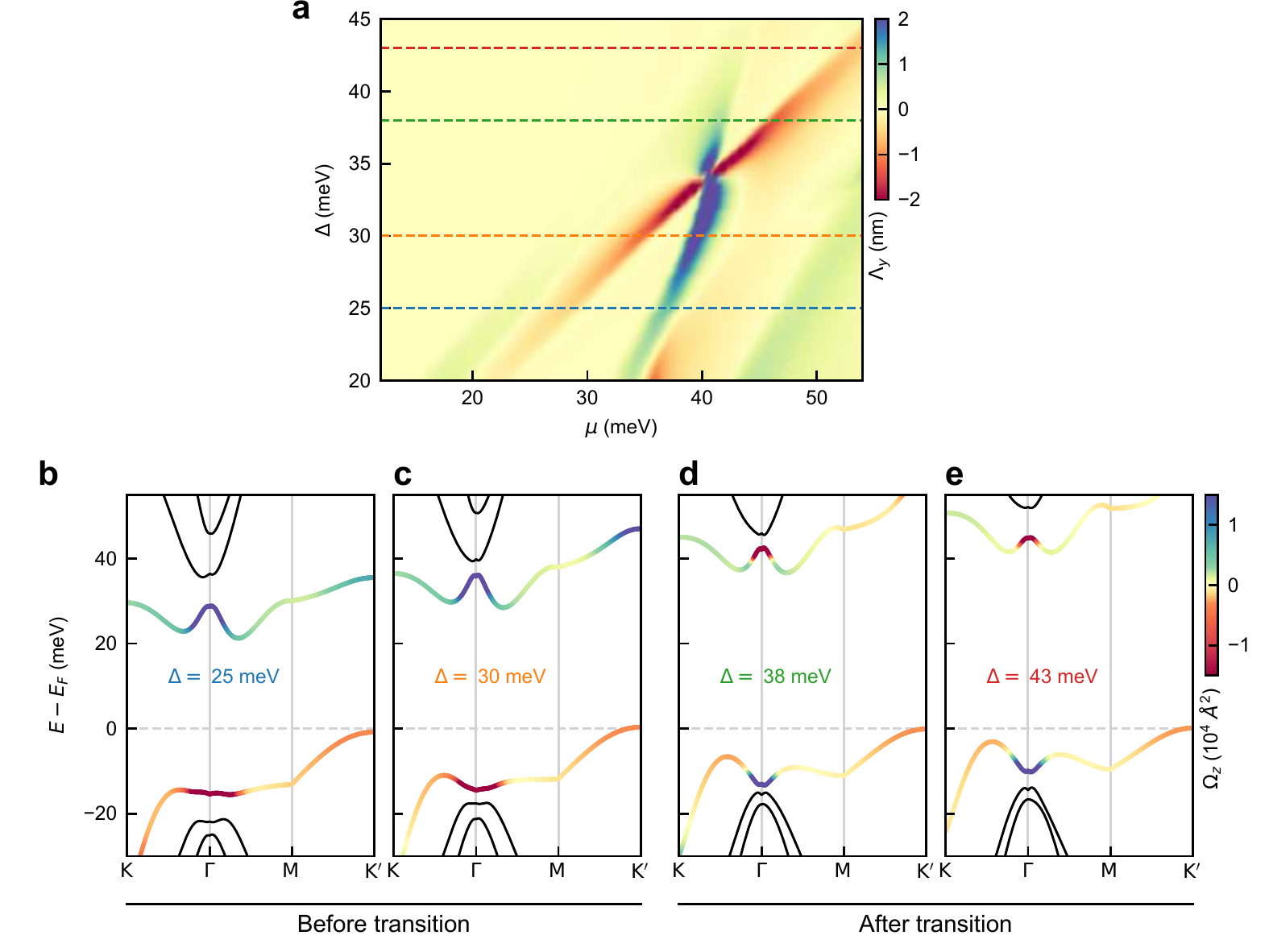}
	\caption{\label{fig:TheoryBCD}{\textbf{Evolution of Berry curvature dipole (BCD). a,}~Dependence of the $y$-component of BCD on the chemical potential ($\mu$) and inter-layer potential ($\Delta$) at the conduction band side. 
			\textbf{b, c,}~Energy dispersion along high symmetry k-paths for $\Delta= 25$~meV~(\textbf{b}) and $\Delta= 30$~meV~(\textbf{c}) before transition. 
			\textbf{d, e,}~Similar energy dispersion for $\Delta= 38$~meV~(\textbf{d}) and $\Delta= 43$~meV~(\textbf{e}) after transition. The color map shows the Berry curvature value for the flat bands.}}
\end{figure*}

 In  Extended Data Fig.~\ref{fig:TheoryBCD}, we have plotted BCD as a function of inter-layer potential ($\Delta$) and chemical potential ($\mu$) at a fixed strain value of $0.1 \%$. 
The butterfly structure signifies the evolution of BCD with the perpendicular electric field. 
The sign reversal of BCD near critical inter-layer potential $\Delta=34$ meV indicates a topological transition. 
The upper panel of Extended Data Fig.~\ref{fig:TheoryBCD} shows the $y$-component of BCD in the conduction band side ($\mu > 0$). 
Before phase-transition, at 25 meV and 30 meV, the red lobe signifies the negative BCD for the first conduction band while the blue lobe represents positive BCD for moir\'e conduction bands. 
The corresponding band dispersions are shown in Extended Data Fig.~\ref{fig:TheoryBCD}b, c.
The overlayed color indicates the Berry curvature hot spots within the band. After the phase-transition, at $\Delta =$ 38 and 43 meV, we observe positive lobe of BCD for the first conduction band while negative lobe for the higher conduction band. 
The corresponding dispersions are shown in Extended Data Fig.~\ref{fig:TheoryBCD}d, e. 
From the color map of Berry curvature, it is evident that in regime-I (before transition), the valence (conduction) band has a negative (positive) Berry curvature hotspot while it gets reversed in regime-II (after transition). 

\section{Metastable states in h-BN encapsulated twisted double bilayer graphene}
\begin{figure*}
	\centering
	\includegraphics[width=16cm]{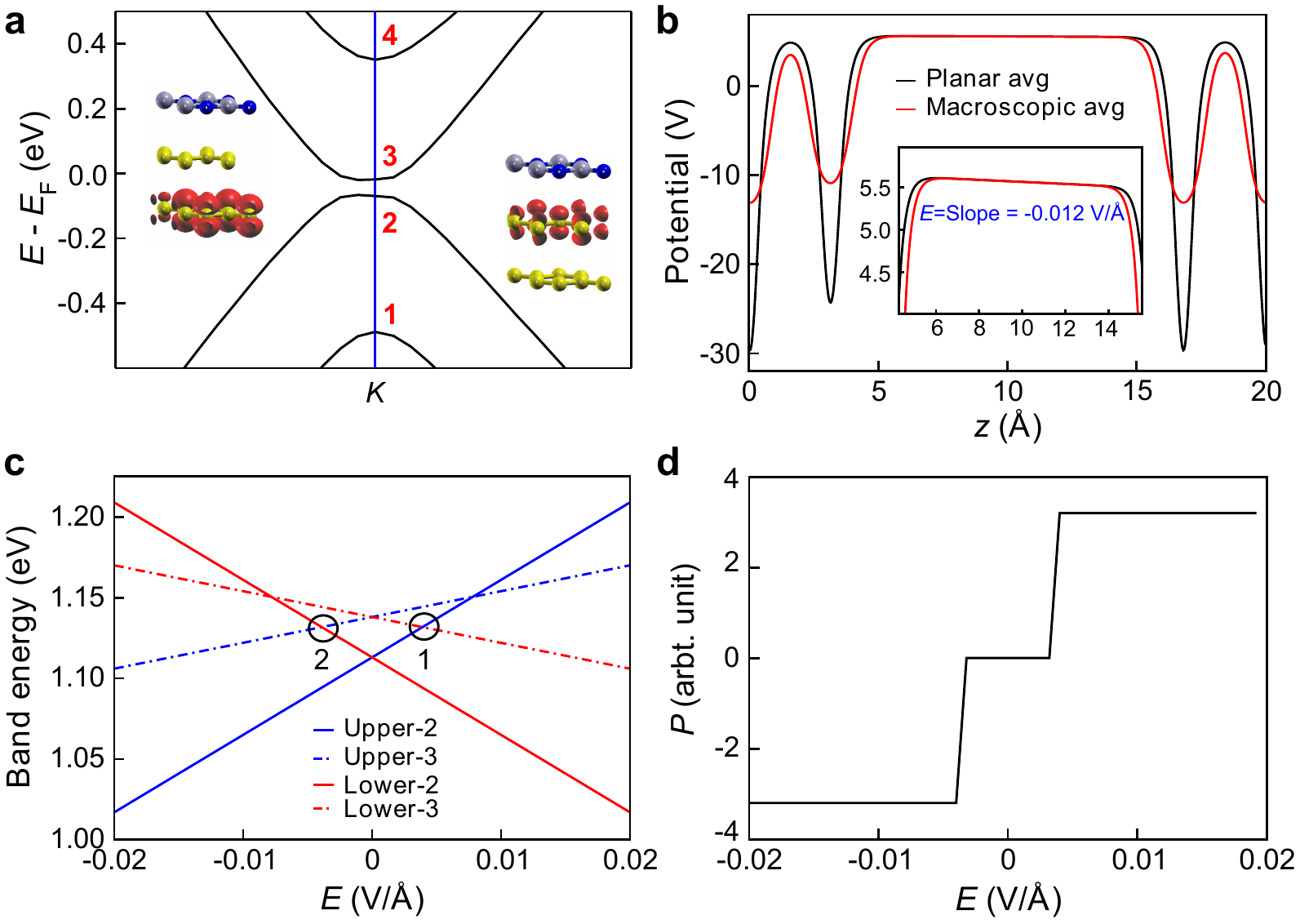}
	\caption{\label{fig:Polarization}{ \textbf{Metastable states and polarization in TDBG.}}  \textbf{a,}~Electronic structure of Gr-Gr-h-BN shows a band gap of 26 meV at \textit{K} point. The inset shows the spatial distribution of the wave functions of bands labelled 2 and 3. 
		\textbf{b,}~Electric field (E) calculated from the slope of the average macroscopic potential of Gr-Gr-h-BN in vacuum. 
		\textbf{c,}~Evolution and crossing of band 2 and 3 at E~$=-0.0039$~V/{\AA} and E~$=0.0039$~V/{\AA} of upper and lower trilayer of h-BN-TDBG-h-BN as a function of electric field using our rigid band model. 
		\textbf{d,}~metastable states for (i) E~$< -0.0039$~V/{\AA} and (ii) E~$> 0.0039$~V/{\AA}, with nonzero polarization in h-BN-TDBG-h-BN that are accessible with the electric field.}
\end{figure*}

To identify the metastable states that possibly cause the observed hysteresis, we first present first-principles density functional theory analysis of a Gr-Gr-h-BN trilayer, which constitutes one of the halves which are twisted by a small angle in the TDBG. 
Interestingly, the outer h-BN monolayer breaks the sublattice-symmetry opening up a small gap at the Fermi level (see Extended Data Fig.~\ref{fig:Polarization}a), accompanied by a weak layer asymmetry in the constitution of frontier electronic states (bands of \textit{p}\textsubscript{z} orbitals) reflecting on an accumulation of a small electronic charge on one of the two graphene monolayers. 
This is expected from the lack of inversion or horizontal reflection symmetries in the trilayer, and our estimate of its polarization (along the $z$-axis) is -0.34 $\mu$C/cm\textsuperscript{2}. 
We analyze effects of perpendicular electric field on the TDBG in two steps, focusing first on the coupling of field with the DBG (twist angle = 0) and then discuss the effect of twist in terms of variation in stacking between the DBGs. 
Using a rigid band model of frontier electronic states of DBG (for details of rigid band model refer to Supplementary Information section XI), we demonstrate that electric field permits switching to two metastable states with opposite polarization (Extended Data Fig.~\ref{fig:Polarization}d) arising from the inversion of frontier electronic states in one of the DBGs (see Extended Data Fig.~\ref{fig:Polarization}c). 
However, these states are symmetry equivalent and would not explain the hysteresis observed in resistances of TDBG, which we argue to arise from two mechanisms: (a) the inhomogeneous doping across the layers in the TDBG due to differences in the top and bottom doping as evident in our calculations of gated TDBG with a twist angle of 21.78\textsuperscript{o}, and (b) switchable ferroelectric polarization associated with redistribution of regions with different stacking sequences (AA versus AB) across the twisted layers as shown in \cite{Yasuda1458}. 

Our calculations of the TDBG (with 21.78\textsuperscript{o}) with a bottom gate reveals the development of polarization upon doping with an estimate of 0.6~\textmu C/cm\textsuperscript{2} (refer to Supplementary Fig.~19a). 
This is comparable to \textit{P} $\sim$ 0.68 \textmu C/cm\textsuperscript{2} of bilayered h-BN \cite{Yasuda1458} and larger than that (\textit{P} $\sim$ -0.18 \textmu C/cm\textsuperscript{2}) arising from the restructuring in terms of stacking sequence \cite{Zheng2020} in heterostructures. 
Thus, the observed hysteresis has possible contributions from inhomogeneous stacking as well as inhomogeneous, layer-dependent carrier concentration.


\clearpage

\begin{center}
	\textbf{\huge Supplementary Information}
\end{center}
\renewcommand{\thesection}{\Roman{section}}
\setcounter{section}{0}
\renewcommand{\thefigure}{S\arabic{figure}}
\captionsetup[figure]{labelfont={bf},name={Supplementary Fig.}}
\setcounter{figure}{0}
\renewcommand{\theHequation}{Sequation.\theequation}
\renewcommand{\theequation}{S\arabic{equation}}
\setcounter{equation}{0}

\section{Moir\'e flat bands in TDBG}
To obtain the electronic band structure and related topological properties of the AB-AB twisted double bilayer graphene (TDBG) with twist angle $1.1^\circ$ we follow the continuum model approach of Bistritzer and MacDonald~\cite{Bistritzer12233}. 
The AB-AB TDBG can be fabricated by placing two AB-stacked bilayer graphene (see the side view in Supplementary Fig.~\ref{fig_1}a) on top of each other and rotate the bilayers with respect to each other. For the band structure calculation, we assume that the upper bilayer ($l=1$) is rotated by an angle $\theta/2$ and the lower bilayer ($l=2$) is rotated by an angle $-\theta/2$ adding to total twist angle $\theta$. 
The smallest continuum Hamiltonian near the ${\bf K}$ valley can be written as~\cite{chebrolu_flat_2019}
\be 
H = 
\begin{pmatrix}
	h_t^++\Delta_t^+ & t_k^{+} & 0 & 0 \\
	{t_k^+}^\dagger & h_b^{+}+\Delta_b^+ & T({\bf r}) & 0 \\
	0 &  T({\bf r})^\dagger & h_t^-+\Delta_t^- &  t_k^-\\
	0 & 0 & {t_k^-}^\dagger  & h_b^{-}+\Delta_b^- \\
\end{pmatrix} .\label{tdbg_ham}
\ee
Here, $+(-)$ sign stands for the $l=1 (2)$ bilayer and $t (b)$ represents the top (bottom) layer of each bilayer. Due to rotation, the Dirac Hamiltonian modifies as $h^{\pm}=R(\mp \theta/2) \hbar v_F {\bf k~\cdot}~\sigma$, where ${\bf k}$ is the crystal momentum near the valley and $\sigma=(\sigma_x, \sigma_y)$ are the Pauli matrices representing the sub-lattice degree of freedom of single-layer graphene. The effect of dimer site potential on each layer of a bilayer has been captured in the Hamiltonian by a parameter $\delta$ as $h_{(t/b)} = 
\hbar v_F \sigma \cdot {\bf k}  + \delta ( \mathbb{1} \mp \sigma_z)/2$. The inter-layer coupling matrix $t_k$ within each bilayer is represented as
\[
\quad t_k= \begin{pmatrix}
	-\hbar v_4 \pi^\dagger & -\hbar v_3 \pi \\ \gamma_1 & -\hbar v_4 \pi^\dagger
\end{pmatrix},
\]
where $\pi=k_x + i k_y$. The strong inter-layer coupling between the two dimer sites is denoted by $\gamma_1$ (shown by solid line in Supplementary Fig.~\ref{fig_1}a), the inter-layer hopping between two non-dimer sites is denoted by $v_3$ (shown by dashed line in Supplementary Fig.~\ref{fig_1}a) and the inter-layer hopping between dimer and non-dimer sites is denoted by $v_4$ (shown by dotted line in Supplementary Fig.~\ref{fig_1}a). 
The various velocities can be calculated from the corresponding hopping amplitudes ($\gamma_i$) using the conversion rule $v_i = \sqrt{3} |\gamma_i| a /(2 \hbar)$ with $a=2.46$ \AA. For our calculations,  we consider the following parameters: the dimer site potential $\delta=15 $ meV, the intra-layer hopping $\gamma_0 = -3.1$ eV, which gives the Fermi velocity, $v_F=10^6$ m/s and $\gamma_1 = 361$ meV. 
The remote hopping amplitudes $\gamma_3$ and $\gamma_4$ are considered to be $283$ meV and $138$ meV, respectively.

\begin{figure*}[t!]
	\centering
	\includegraphics[width=0.85\linewidth]{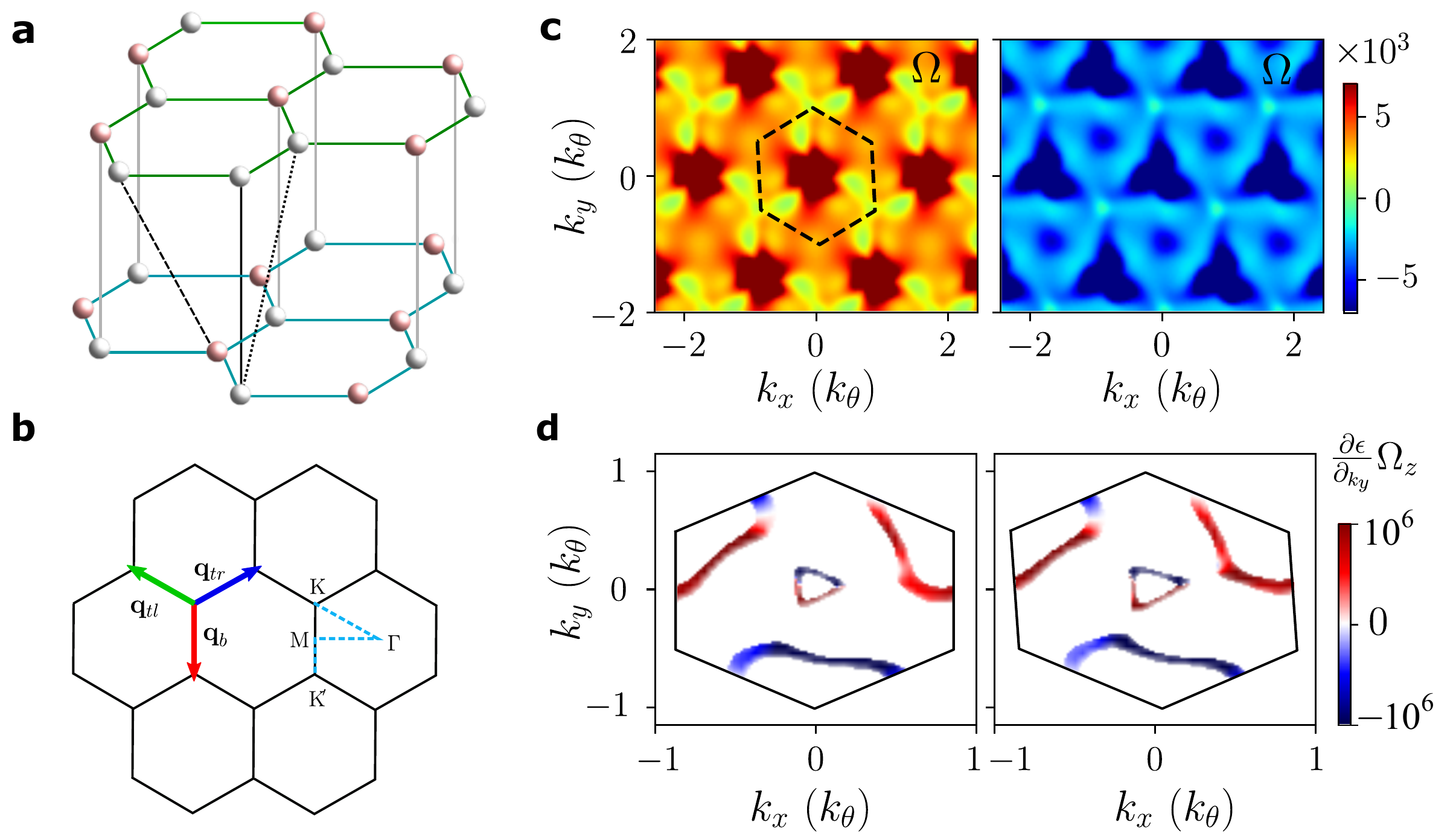}
	\caption{\footnotesize \textbf{a,} The side view of the Bernal stacked (AB-stacked) bilayer graphene. The hopping paths corresponding to $\gamma_1$, $\gamma_3$ and $\gamma_4$ are shown by black solid, dashed and dotted lines. 
		\textbf{b,} Schematic of hexagonal moir\'{e} BZ. Three nearest neighbor connecting vectors, ${\bf q}_{b}$, ${\bf q}_{tr}$ and ${\bf q}_{tl}$ are shown by red, blue and green arrows respectively. The high symmetry points in the moir\'e BZ are indicated by ${\rm K}, {\rm \Gamma}, {\rm M}$ and ${\rm K}'$ along which we plot the band dispersion (shown by cyan dashed line). 
		\textbf{c,} The Berry curvature (in units of \AA$^2$) distribution in the moir\'e BZ for conduction (left) and valence (right) bands in presence of $\epsilon=$ 0.1$\%$ strain. The $k_x$ and $k_y$ axes are normalized with $k_\theta=8\pi/(3a) \sin(\theta/2)$ with $a=2.46~$\AA. 
		\textbf{d,} The distribution of the Berry curvature dipole integrand of Eq.~\eqref{D_cd} [$\frac{\partial \epsilon}{\partial k_y} \Omega_z$] over the Fermi surface in units of meV$\cdot$\AA$^3$, in the isolated conduction flat band, without 
		strain (left panel) and with strain (right panel) applied along the zigzag edge of graphene. }
	\label{fig_1}
\end{figure*}

The moir\'e coupling between the twisted bilayers (coupling between the two adjacent rotated layers) in Eq.~\eqref{tdbg_ham} can be expressed as,  
$T({\bf r})=\sum_{j=b, tl, tr}T_{{\bf q}_j} e^{-i{\bf q}_j \cdot {\bf r}}$. The hopping paths are given by the three vectors ${\bf q}_b= \frac{8\pi}{3a} \sin \frac{\theta}{2}(0, -1)$, ${\bf q}_{tr}= \frac{8\pi}{3 a} \sin \frac{\theta}{2}(\frac{\sqrt{3}}{2}, \frac{1}{2}$) and ${\bf q}_{tl}= \frac{8\pi}{3 a} \sin \frac{\theta}{2}(-\frac{\sqrt{3}}{2}, 1/2)$ and the hopping matrices are given by, 
\be 
T_b = 
\begin{pmatrix}
	\omega' & \omega \\
	\omega & \omega'
\end{pmatrix};
\hspace{0.2 cm} T_{tr,tl} = 
\begin{pmatrix}
	\omega' & \omega e^{\mp i 2\pi/3} \\
	\omega e^{\mp i 2\pi/3} & \omega'
\end{pmatrix}.
\ee
The diagonal ($\omega'$) and off-diagonal ($\omega$) hopping strengths have been considered to be unequal due to the out-of-plane corrugation effect and chosen in the scale $79$~meV and $106$~meV, respectively. The tunability of the electronic band structure of the TDBG due to perpendicular electric field is included in the model Hamiltonian by means of inter-layer potential difference parameter $\Delta$. To model the effective electric field as a constant gradient in potential, we use $\Delta^{-}_b$=$-\Delta^{+}_t$=$\frac{3}{2} \Delta$ and $\Delta^{-}_t$=$-\Delta^{+}_b$=$\frac{1}{2} \Delta$.

\section{Impact of strain on electronic structure of TDBG}
Equation~\eqref{tdbg_ham} and the corresponding band structure explain the measured resistance as a function of charge density and displacement field reasonably well. However, it can not account for the measured nonlinear Hall voltage due to the presence of $C_3$ symmetry in the continuum model. So to break the $C_3$ symmetry we include the effect of uniaxial strain quantified as~\cite{He2020}
\be \label{strain}
{\mathcal E} = \epsilon
\begin{pmatrix}
	-\cos^2 \phi + \nu \sin^2 \phi & -(1+ \nu) \sin \phi \cos \phi \\
	-(1+ \nu) \sin \phi \cos \phi & -\sin^2 \phi + \nu \cos^2 \phi  
\end{pmatrix},
\ee
which breaks the $C_3$ symmetry~\cite{Bi2019, He2020, zhang_giant_2020, hu_nonlinear_2020}. In Eq.~\eqref{strain} $\epsilon$ is the strength of strain, $\nu$ is the Poisson ratio ($\sim 0.16$ for graphene) and $\phi$ is the strain angle with respect to zigzag direction of graphene. The strain matrix has two fold impact on the model Hamiltonian---i) The Dirac points of bilayer gets shifted to {\bf D}$= (1-{\mathcal E}^T)${\bf K}$ - ${\bf A} from ${\bf K}$ with ${\bf A}= \frac{\beta}{d} (\epsilon_{xx}-\epsilon_{yy}, -2\epsilon_{xy})~(\mathrm{where}~\beta=1.57~\mathrm{and}~d=1.42~$\AA~) being the strain induced effective gauge field. As a result the moir\'e coupling vectors and the hopping matrices get modified. ii) The Dirac Hamiltonian itself modifies to $\hbar v_F \hat{R}_{\frac{\theta}{2} }[(1+{\mathcal E}^T){\bf k} \cdot \sigma]$. In our calculation we extend the strain implementation of Ref.~\cite{He2020} for twisted bilayer graphene to the TDBG where the hetero strain is applied solely on the lower bilayer. 
\begin{figure*}[t!]
	\centering
	\includegraphics[width=0.7\linewidth]{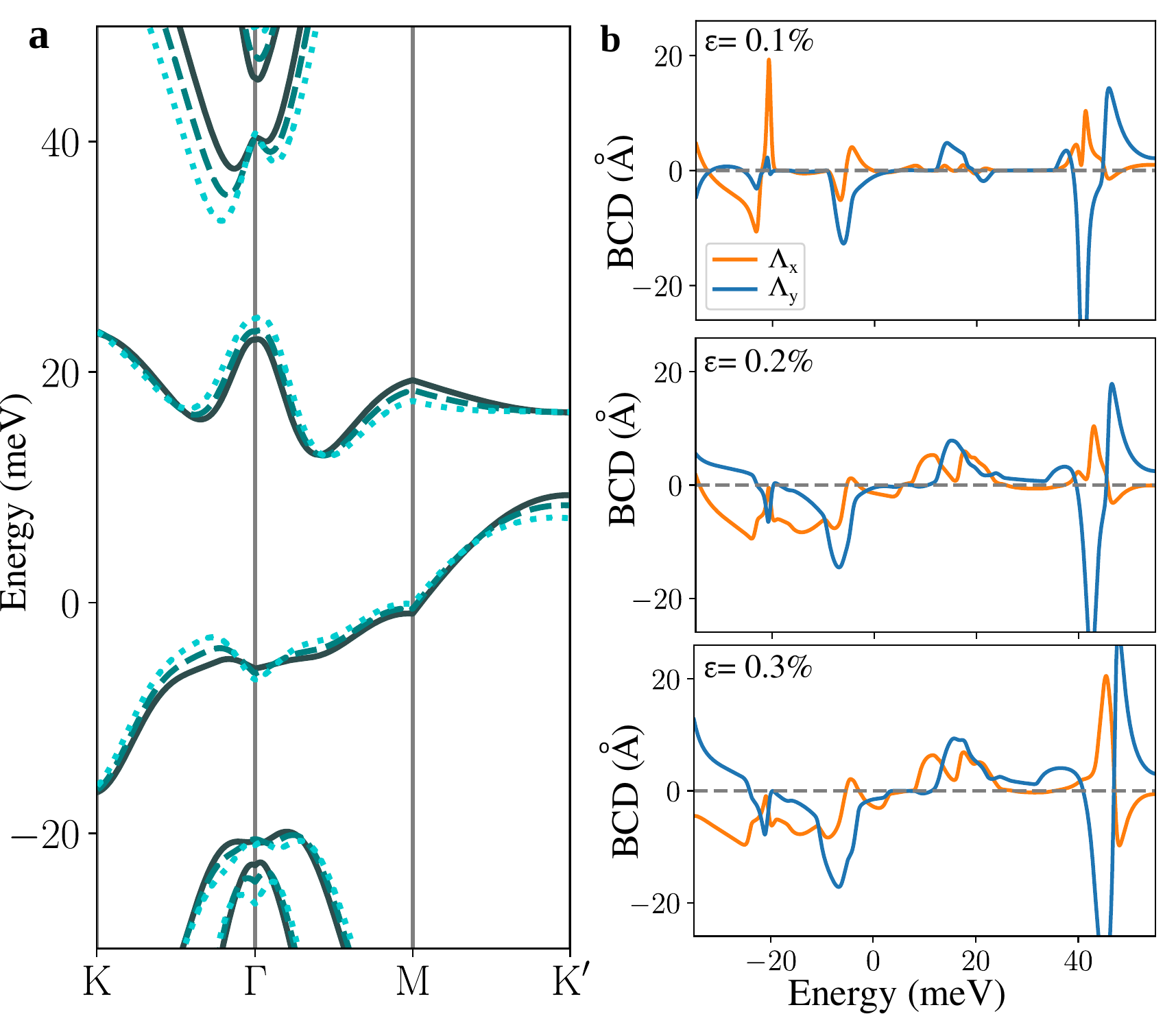}
	\caption{\footnotesize \textbf{a,} Solid, dashed and dotted lines represent band structure with $\epsilon=$ 0.1$\%$, 0.2$\%$ and 0.3$\%$ strain respectively for $\Delta=$ 11 meV. 
		\textbf{b,} In top, middle and bottom panel, the $x$ and $y$ components of Berry curvature dipole are shown for strain stength $\epsilon=$ 0.1$\%$, 0.2$\%$ and 0.3$\%$, respectively. }
	\label{fig_3}
\end{figure*}

The breaking of $C_3$ symmetry after strain implementation in the Hamiltonian is highlighted in Fig.~1b, c of the main manuscript where the Berry curvature ($\Omega$) of the valence band is plotted along three paths arranged in 120$^\circ$ angle as  shown by the blue, orange and green arrows in Fig.~1a of the main manuscript. The Berry curvature is calculated using the periodic part of the Bloch wave-function, $H |u_n \rangle = E_n  |u_n \rangle$, as
\be 
\Omega^n \equiv \Omega^n_z = -2 {\rm Im} \sum_{m \neq n  } \dfrac{ \langle  u_n |\partial_{k_x} H | u_m \rangle \langle  u_m |\partial_{k_y} H | u_n\rangle}{(E_n - E_m)^2} ~.
\ee

In Supplementary Fig.~\ref{fig_1}c we have shown the Berry curvature distribution for the conduction (left column) and valence (right column) bands over the moir\'{e} BZ. Due to the broken $C_3$ symmetry, the distribution of Berry curvature in the ${\bf k}$-space becomes non-symmetric. Such non-symmetric Berry curvature in the moir\'e Brillouin zone (mBZ) recently has been shown to cause nonlinear Hall effect which is quantified by Berry curvature dipole (BCD) defined as
\be \label{D_cd}
D_{cd} = \sum_{n,\xi,g_s} \int_{\rm mBZ}  \dfrac{d{\bm k}}{(2\pi)^2} \Omega_d^n \frac{\partial \epsilon^n_{\bm k}}{\hbar \partial k_c} \frac{\partial f(\epsilon^n_{\bm k})}{\partial \epsilon^n_{\bm k}}.
\ee
Here, $n$ is the band index, $\xi$ is the valley index, $g_s$ is the spin index and $f_0$ is the Fermi-Dirac distribution function. For a two dimensional system (TDBG in our case) the Berry curvature acts as pseudo-scalar and has only $z$-component. So depending on the direction of velocity, $\frac{\partial \epsilon^n}{\partial k_c}$, the BCD has only two components $D_{xz} \equiv \Lambda_x \equiv {\rm BCD}_x$ and $D_{yz} \equiv \Lambda_y \equiv {\rm BCD}_y$. The distribution of the Berry curvature dipole kernel, $\frac{\partial \epsilon}{\partial k_y}\Omega_z$ of the flat conduction band is highlighted in Supplementary Fig.~\ref{fig_1}d where the left panel shows the kernel in absence of strain and the right panel shows the kernel in presence of $0.1\%$ strain. The variation of BCD with the strain strength is shown in Supplementary Fig.~\ref{fig_3} where three different magnitude of strain: $0.1\%$, $0.2\%$ and $0.3\%$ are considered.

\begin{figure*}[t!]
	\centering
	\includegraphics[width=\linewidth]{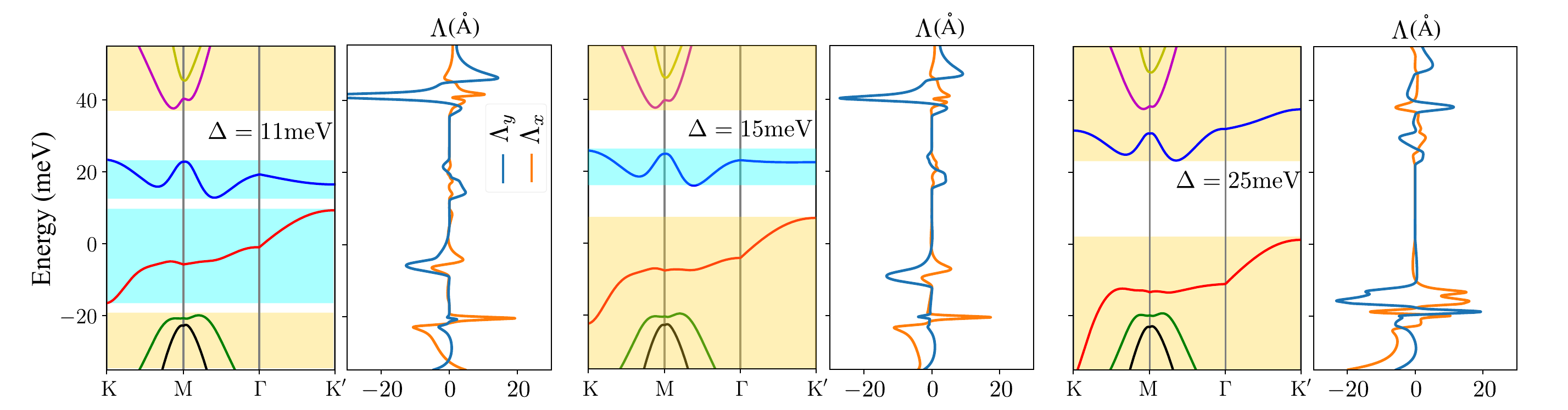}
	\caption{\footnotesize The evolution of band structure and Berry curvature dipole in presence of 0.1$\%$ strain for three different $\Delta$ values: $\Delta=$ 11, 15 and 25 meV. The left most panel shows the BCD when CNP gap as well as both the moir\'{e} gaps are present. The middle one highlights the scenario when the valence band side moir\'{e} gap is closed. The right most panel reveals the BCD when only the CNP gap is present.}
	\label{fig_2}
\end{figure*}

\section{Variation of band structure and Berry curvature dipole with electric field in presence of strain}\label{sec3}

\par One of the experimental advantages of TDBG in comparison to the TBG counterpart is that one can tune the electronic band dispersion and hence related physical properties by an external electric field. This electric field tunability is captured in our Hamiltonian, Eq.~\ref{tdbg_ham}, by the $\Delta$ term. In this section, we discuss how the band structure and Berry curvature dipole evolve with the variation in external electric field in presence of strain. This is shown in Supplementary Fig.~\ref{fig_2} (with $0.1\%$ strain) and we find that the evolution of the band structure is qualitatively consistent with our experimental findings. In absence of an external electric field, i.e. $\Delta=0$ meV, the strained flat bands overlap and promote a metallic state. At the same time the flat bands are separated from the higher moir\'{e} bands by a finite energy gap, 
namely moir\'{e} gap.

Application of finite external electric field pushes the flat bands away from each other. However, a gap at charge neutrality point (CNP) appears only after certain threshold electric field and in our model calculations we find it to be $\sim$ 9.5 meV.  At $\Delta= 11$ meV, the CNP gap as well as both the moir\'{e} gaps exist simultaneously. The corresponding band structure and BCD are shown in the left most column of Supplementary Fig.~\ref{fig_2}. Gradual increment of $\Delta$ enhances the CNP gap and decreases the magnitude of both the moir\'{e} gaps. At $\Delta \sim$ 13 meV the valence flat band merges with higher moir\'{e} bands while the conduction moir\'{e} gap remains finite. This scenario of band structure and corresponding BCD are indicated in the middle column of Supplementary Fig.~\ref{fig_2}.  Further increment in $\Delta$ eventually closes the moir\'{e} gap at conduction band side and only the CNP gap persists. The right most column of Supplementary Fig.~\ref{fig_2} shows the band structure and BCD at $\Delta= 25$ meV.

\section{Sign reversal of Berry curvature dipole across the topological phase transition} 

The electronic wave-functions in TDBG are known to be rich in topological aspects. It has been predicted that tuning the strength of the electric field or twist angle can give rise to a topological phase-transition in this system. However, to the best of our knowledge, topological phase transition in TBG/TDBG has not been experimentally demonstrated. In our experiment we observe a topological phase-transition by tuning the electric field, which manifests as the sign reversal of BCD. The non-trivial topology of the electronic wave-functions in TDBG is characterized by the valley Chern number $({\mathcal C}_v)$~\cite{koshino_band_2019-1}. For an isolated band, the Chern number in each valley can be calculated by integrating the Berry curvature in the mBZ as
\be 
{\mathcal C}_v^n = \dfrac{1}{2\pi} \int_{\rm mBZ}  d{\bf k} \Omega^n .
\ee
Due to the presence of time reversal symmetry, the Chern numbers for $K$ and $K^\prime$ valley are equal and opposite which makes the total (adding the two valleys) Chern number, for a particular band, zero. So to distinguish different topological phases we define a topological invariant (Z) as
Z = |($C_K$ -$C_{K^\prime}$)/2|, where $C_K~\mathrm{and}~C_{K^\prime}$ are the Chern numbers for K and K' valley, respectively. A topological phase transition is identified with a change in the associated Z index, $ \Delta Z$. In presence of $\epsilon=$ 0.1 $\%$ strain we find a topological phase-transition near $\Delta \sim 34$ meV in our calculation. We emphasize that both the phases across the topological transition are robust in a large window of $\Delta$.

The band dispersion, Chern number and Berry curvature dipole in these two distinct topological phases are shown in Fig.~3e-h of the main manuscript. The bands plotted in Fig.~3f  indicate the phase before the transition ($\Delta=$ 25 meV) for the $K$ valley. The corresponding Chern numbers of the first conduction and first valence bands are 2 and -2, respectively resulting in Z$=2$ for both the bands. For $\Delta=$ 38 meV, the band dispersion after the phase-transition, is plotted in Fig.~3e. The calculated Chern numbers for this phase are 0 and 1 for the first conduction and first valence bands, respectively. This results in Z$=0$ for the first conduction band and Z$=1$ for the first valence band. Therefore the phase transition in conduction band is conveyed through $\Delta Z=2$ and in the valence band through $\Delta Z=1$. Note that some recent literature reports the Chern numbers of the flat bands can be tunable upto $\pm$3 with variation in twist angle and electric field for the unstrained moir\'{e} systems~\cite{xiang_phase_2021, Koshino2019, mohan_trigonal_2021}. Interestingly the  phase-transition near $\Delta=34$ meV is also evident from the band structure evolution across the critical point. Focusing on the conduction band side, we find that as we gradually increase the electric field from $\Delta=25$ meV, the first conduction band gets closer to the higher conduction bands. However, it starts to move away from the higher band as we cross the critical point. Remarkably, we find that the BCD changes its sign across this phase transition which is also highlighted in Fig.~3g, h of the main manuscript. The similar BCD peaks for two different phases are shown with arrows of same color.  We emphasize here the exact values of Chern numbers crucially depends on the chosen parameters. However, the topological transition which is associated with finite value of $\Delta Z$ for each flat band can be seen for a broad range of parameters. 
\begin{figure*}[t!]
	\centering
	\includegraphics[width=0.75\linewidth]{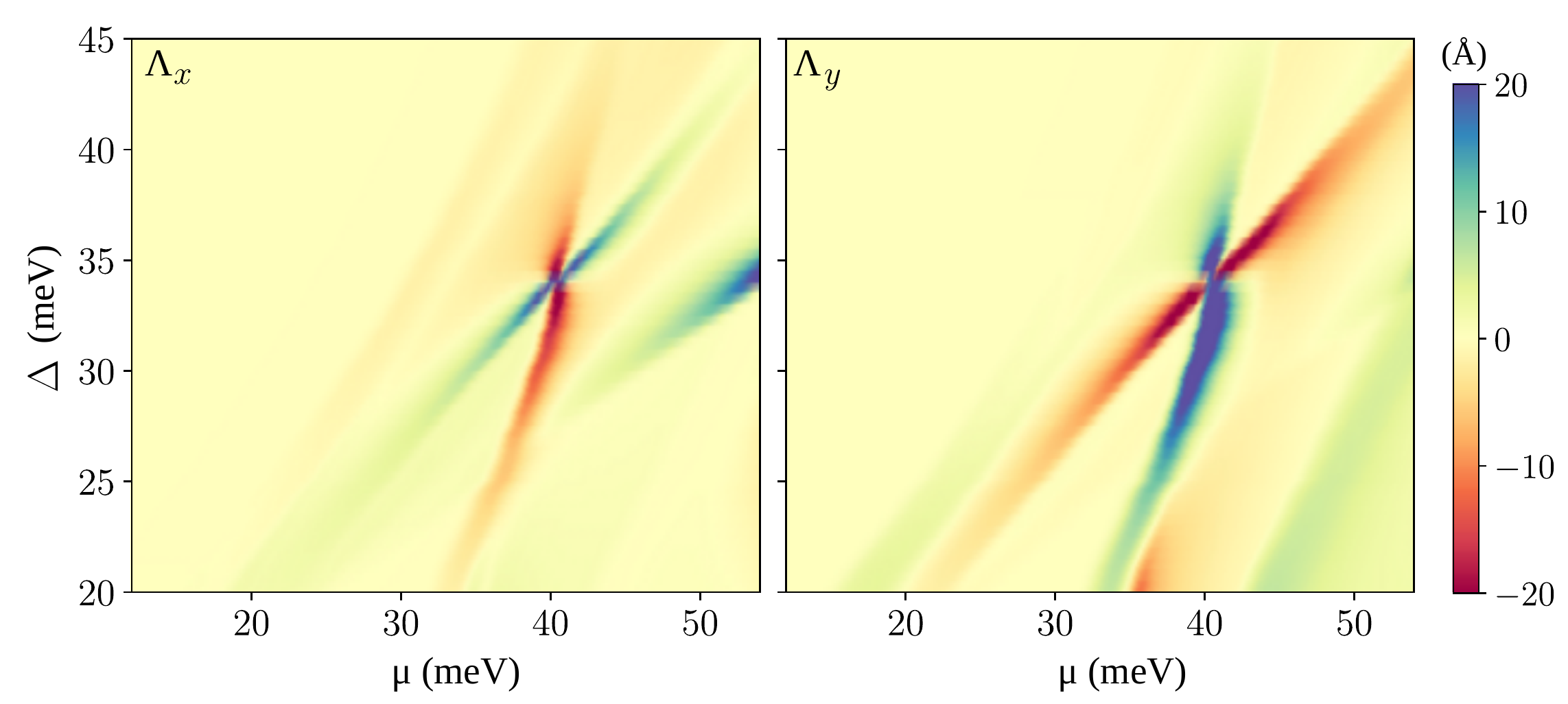}
	\caption{\footnotesize Dependence of the Berry curvature dipole on the chemical potential ($\mu$) and perpendicular electric field ($\Delta$)  at the conduction band side. The butterfly like structure near $\Delta=34$ meV indicates a perpendicular electric field induced topological phase-transition.}
	\label{fig_4}
\end{figure*}

\par To emphasize the topological phase-transition more comprehensively we have plotted the $x$- and $y$-components of BCD as a function of perpendicular electric field ($\Delta$) and chemical potential ($\mu$) in Supplementary Fig.~\ref{fig_4} in presence of 0.1$\%$ strain along zigzag direction ($\phi=0$). A clear sign of perpendicular electric field induced phase-transition can be realized from the butterfly like structure near $\Delta=34$ meV. Before the phase-transition ($\Delta< 34$ meV) the $x$-component of BCD of the first conduction band has positive sign (shown by the blue lobe) and the second conduction band has negative sign (shown by the red lobe). However, after the phase-transition this trend gets reversed as the first conduction band possesses negative BCD$_x$ (shown by the red lobe) and the second conduction band possesses positive $\Lambda_x$ (shown by the blue lobe). Apart from the fact that the sign of the $y$-components of the BCD is opposite to the corresponding $x$-component, $\Lambda_y$ also shows a similar behavior. This abrupt change in the sign of BCD near $\Delta \sim$ 34 meV indicates a topological transition. We emphasize here that we have considered $\mu=0$ to reside in the maxima of the valence band and the shaded region near $\mu~\sim$ 50 meV is contribution of the further higher conduction bands to the BCD.

\section{Origin of BCD in graphene based moir\'e materials}
In this section we explore the origin of large BCD in graphene based moir\'e systems, in vicinity of the charge neutrality point. For that we calculate Berry curvature dipole for a low energy tilted massive Dirac model and extend it to strained two band model of twisted bilayer graphene~\cite{mannai_twistronics_2021}. The nonlinear Hall conductivity for a tilted massive Dirac Hamiltonian of the form 
\be 
{\mathcal H} = \hbar t  k_x \sigma_0 + \hbar v k_x \sigma_x + \hbar v k_y \sigma_y + m\sigma_z
\ee
can be written as $\sigma_{yxx} = \frac{e^3 \tau }{\hbar} {\rm BCD}_x$ in the limit $\omega \tau \ll 1$, which is valid for transport experiments. Up to  linear order in the tilt value, the  $x$-component of the BCD at chemical potential $\mu$ is given by
\be \label{D_x}
{\rm BCD}_x =  \dfrac{3m \hbar t}{8 \pi \mu^2} \left( 1 - \dfrac{m^2}{\mu^2}\right).
\ee
From this expression we infer that within the constant scattering time approximation the BCD is proportional to the tilt and the BCD peak in $\mu$-axis is determined by the tilt and Berry curvature hotspot. 

The low energy mode for strained twisted bilayer graphene given, in vicinity of the charge neutrality point, is given 
by~\cite{mannai_twistronics_2021}
\bea  \label{ham}
{\mathcal H} = - \dfrac{\hbar}{1 + 6 \alpha^2} \psi_0^\dagger \Big[v_{0x} k_x + v_{0y} k_y + \xi \sigma_x v_x k_x + \sigma_y v_y k_y 
+ \xi \sigma_{x} v_{xy} k_y + \sigma_y v_{yx} k_x + m \sigma_z
\Big] \psi_0.
\eea 
Here, $\alpha=w/(\hbar v_F k_\theta)$ with $k_\theta=8 \pi/(3a) \sin (\theta/2)$ and the various velocities are modified by strain as shown in Ref.~\cite{mannai_twistronics_2021}. In Eq.~\eqref{ham} while $v_{0x}, v_x, v_y$ are determined by the off-diagonal components of strain matrix, the other components $v_{0y}, v_{xy}, v_{yx}$ are determined by the diagonal components of strain matrix~\cite{mannai_twistronics_2021}. For a rough estimation, we will consider shear strain (diagonal components of strain are zero) where the tilt velocity is given by 
\be 
v_{0x} = - \xi v_F \dfrac{ \alpha^2}{1 +6 \alpha^2}  \dfrac{16 \pi}{a k_\theta} \epsilon_{xy}.
\ee
Using this tilt value in Eq.~\eqref{D_x}, we have  
\be
{\rm BCD}_x \propto t \propto v_{0x} \propto k_\theta^{-3} \epsilon_{xy}. 
\ee
This simple estimate indicates that the BCD is larger for systems with i) large moir\'e lengthscale, and ii) with a large strain.

\section{Scaling of nonlinear Hall Effect}

In this section we describe the scaling law of the experimentally measured quantity $\frac{V_y^{N}}{(V_{x}^L)^2}$ which is related to the theoretically calculated nonlinear Hall conductivity $\sigma_{yxx}$ and Drude conductivity $\sigma_{xx}$ as
\be \label{ratio}
\dfrac{V_y^{N}}{(V_{x}^L)^2} = \dfrac{\sigma_{yxx}}{\sigma_{xx}}~.
\ee
The non-linear Hall conductivity can originate from three different sources: i) the Berry curvature dipole, ii) side-jump scattering and iii) skew-scattering (also called anti-symmetric scattering). 
For a tilted massive Dirac model Hamiltonian it can be shown that~\cite{du_disorder-induced_2019} the Berry curvature dipole and the side-jump contributions to $\sigma_{yxx}$ are inversely proportional to the impurity concentration making the ratio in Eq.~\eqref{ratio} to be scattering independent. However, the skew-scattering part has different impurity concentration dependence. Accounting for all the terms, we can write a general scaling relation, following Du et al.~\cite{du_disorder-induced_2019}, 
\be
\dfrac{V_y^{N}}{(V_{x}^L)^2} = {\mathcal C}^{in} + \sum_i {\mathcal C}_i^{sj} \dfrac{\rho_i}{\rho_{xx}} + \sum_{i,j} {\mathcal C}_{ij}^{sk1} \dfrac{\rho_i \rho_j}{\rho_{xx}^2} + \sum_i {\mathcal C}_i^{sk2} \dfrac{\rho_i}{\rho_{xx}^2} ~.
\ee
Here, $i,j$ represent different source of scattering, the superscripts {\it in}, {\it sj} and {\it sk} stands for Berry curvature dipole, side-jump and skew-scattering contributions respectively. Considering only two sources of scattering, the static (impurities) and dynamic (phonon), we can write the above equation as follows
\be  \label{sc}
\dfrac{V_y^{N}}{(V_{x}^L)^2}=\dfrac{1}{\rho_{xx}^2}\left({\mathcal C}_1 \rho_{xx0} +{\mathcal C}_2 \rho_{xx0}^2+  {\mathcal C}_3 \rho_{xx0} \rho_{xxT} +   {\mathcal C}_4 \rho_{xxT}^2 \right)~.
\ee
Here, $\rho_{xx0}$ is the residual resistivity and $\rho_{xxT}=\rho_{xx}-\rho_{xx0}$ is the dynamical resistivity.
The new parameter set in Eq.~\eqref{sc} can be obtained from the old one as
\bea
{\mathcal C}_1&=&{\mathcal C}^{sk2}_{0} ;{\mathcal C}_2={\mathcal C}^{in} +{\mathcal C}^{sj}_0+ {\mathcal C}^{sk1}_{00}, \\
{\mathcal C}_3&=&2{\mathcal C}^{in} +{\mathcal C}^{sj}_0+{\mathcal C}^{sj}_1+ {\mathcal C}^{sk2}_{01}, \\
{\mathcal C}_4&=&{\mathcal C}^{in} +{\mathcal C}^{sj}_1+ {\mathcal C}^{sk1}_{11} .
\eea
For finite temperature, using $\rho_{xxT}=\rho_{xx} - \rho_{xx0}$, we can write the scaling law in terms of the conductivities as
\bea \nn
\dfrac{V_y^{N}}{(V_{x}^L)^2} -{\mathcal C}_1 \sigma_{xx0}^{-1} \sigma_{xx}^2 =\left({\mathcal C}_2 + {\mathcal C}_4 - {\mathcal C}_3 \right) \sigma_{xx0}^{-2} \sigma_{xx}^2\\ \label{ratio_2}
+ \left( {\mathcal C}_3 - 2{\mathcal C}_4\right) \sigma_{xx0}^{-1} \sigma_{xx}+   {\mathcal C}_4.
\eea
For experimental fitting we use a more simplified scaling law which reads as
\be 
\dfrac{V_y^{N}}{(V_{x}^L)^2} = A \sigma_{xx}^2  +B~,
\ee
where $A$ represents the slope and $B$ is the intercept. From Eq.~\eqref{ratio_2} it is evident that the slope $A$ does not include any Berry curvature dipole contribution and is solely determined by the skew-scattering and side-jump mechanism.
The intercept $B$ contains information of BCD.

\section{NLH voltage from other TDBG devices}

In Supplementary Fig.~\ref{fig:nlh_second_dev}a and Supplementary Fig.~\ref{fig:nlh_second_dev}b, we show the dependence of longitudinal resistance and nonlinear Hall voltage, respectively, on filling factor $\nu$ and displacement field $D/\epsilon_0$. 
The data is taken from device-1 presented in main manuscript having a twist angle of 1.1$^\circ$, using a different set of probes.
Supplementary Fig.~\ref{fig:nlh_second_dev}c and Supplementary Fig.~\ref{fig:nlh_second_dev}d shows the dependence of longitudinal resistance and nonlinear Hall voltage, respectively, on $\nu$ and $D/\epsilon_0$ from a different TDBG device (device-2) with a twist angle of 1.26$^\circ$. 
The additional vertical feature of high resistance in Supplementary Fig.~\ref{fig:nlh_second_dev}c close to $n=2.9 \cross 10^{12}$~cm$^{-2}$ can be attributed to \moire gap from a regime in device-2 having a slightly different twist angle of 1.09$^\circ$.
Twist angle angle variation in same device is a signature of strain that breaks $C_3$ symmetry and give rise to BCD.
In all the cases, the characteristic cross and halo feature that is present in the longitudinal resistance color plots (in Supplementary Fig.~\ref{fig:nlh_second_dev}a, Supplementary Fig.~\ref{fig:nlh_second_dev}c herein and also in Fig.~2a of main manuscript) is absent in the NLH voltage color plots (in Supplementary Fig.~\ref{fig:nlh_second_dev}b, Supplementary Fig~\ref{fig:nlh_second_dev}d herein and also in Fig.~2b of main manuscript). 
This shows the universality of NLH voltage across multiple TDBG devices and that the NLH signal only persists close to either the CNP or the \moire gaps, as discussed in the main manuscript. 

\begin{figure*}
	\centering
	\includegraphics[width=15.5cm]{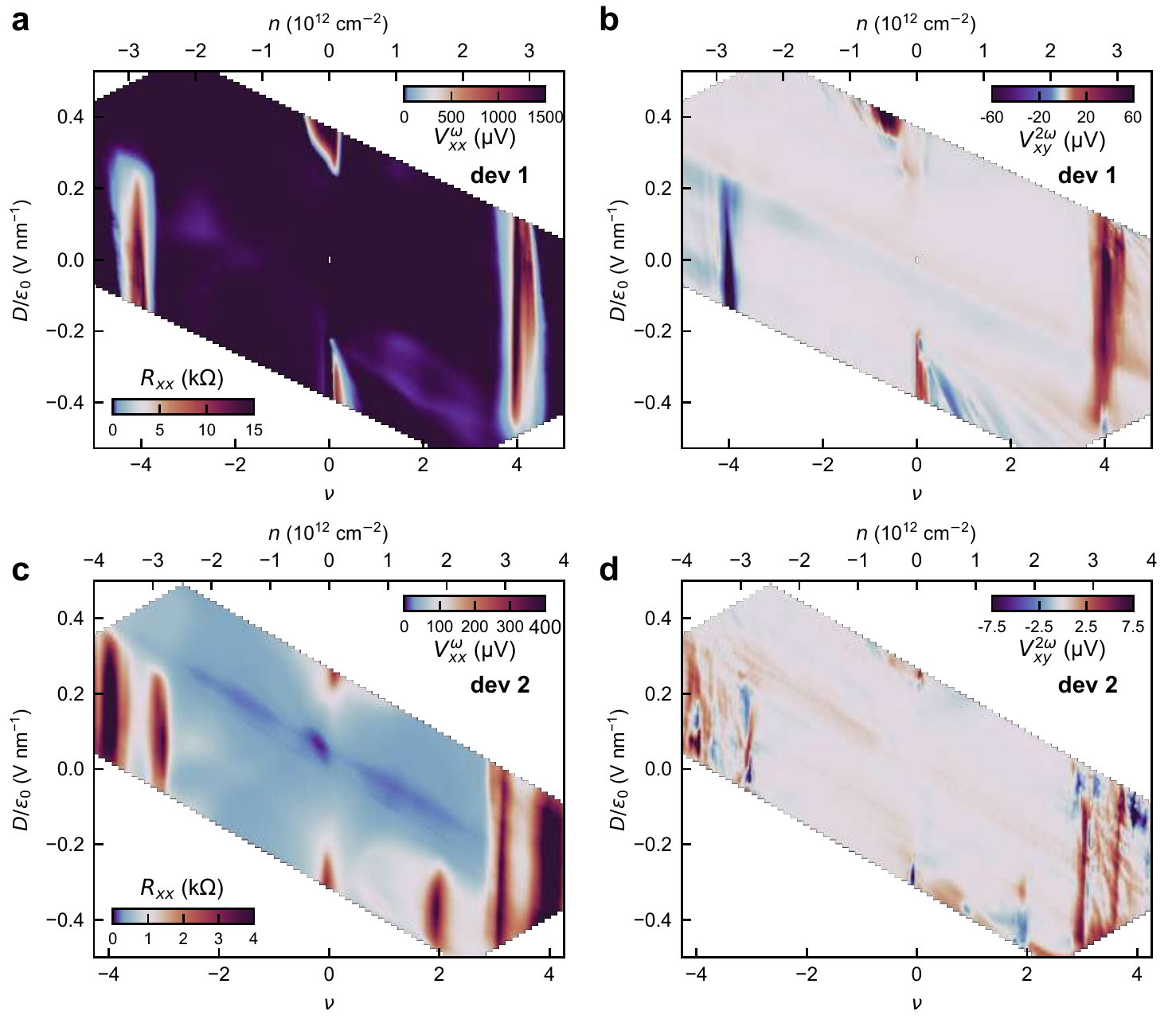}
	\caption{ \label{fig:nlh_second_dev} {\footnotesize \textbf{NLH voltage in other TDBGs.}}
		\textbf{a, b,}~Longitudinal voltage (\Vxxw) (\textbf{a}) and nonlinear Hall voltage (\Vxytw) (\textbf{b}) as a function of filling factor ($\nu$) and perpendicular electric displacement field ($D/\epsilon_0$) at a temperature of 1.5~K. The voltages are measured using a different set of probes than that in main manuscript for the 1.1$^\circ$ device. \textbf{c, d,}~Longitudinal voltage (\Vxxw) (\textbf{c}) and Nonlinear Hall voltage (\Vxytw) (\textbf{d}) as a function of $\nu$ and $D/\epsilon_0$ at a temperature of 10~mK for another TDBG device with a twist angle of 1.26$^\circ$. The top x-axis indicates the charge density ($n$). The color bar provided in bottom left of \textbf{a} and \textbf{c} indicates the corresponding values of the longitudinal resistance ($R_{xx}$) measured using 4-probe method. All the measurements are performed using a constant current $I= 100$~nA sent with frequency $\omega=177$~Hz.}
\end{figure*}

\section{Characterization of NLH voltage}
\subsection{Quadratic nature}
In Supplementary Fig.~\ref{fig:quadratic}a, we show the quadratic scaling of the NLH voltage (\Vxytw) with current for the same filling $\nu=0.125$ used in Fig.~3a and 3b of main manuscript.
In Fig.~\ref{fig:quadratic}b shows the linear dependence of \Vxytw{} on square of the longitudinal voltage (\Vxxw{}).
A linear behavior establishes quadratic scaling of NLH voltage, that we measure, with current.
Supplementary Fig.~\ref{fig:quadratic}a and Supplementary Fig.~\ref{fig:quadratic}b together shows that the quadratic nature persists in the regime of displacement field we use to extract BCD in Fig~3b of main manuscript.
Supplementary Fig.~\ref{fig:quadratic}c shows the quadratic scaling of the NLH voltage that corresponds to a filling $\nu=4.288$ close to the electron-side \moire gap.
We see departure from linear behavior in Supplementary Fig.~\ref{fig:quadratic}d for few displacement fields towards high values of \Vxxw$^2$. 
This is outside the scope of our present study.  
\begin{figure*}
	\centering
	\includegraphics[width=15.5cm]{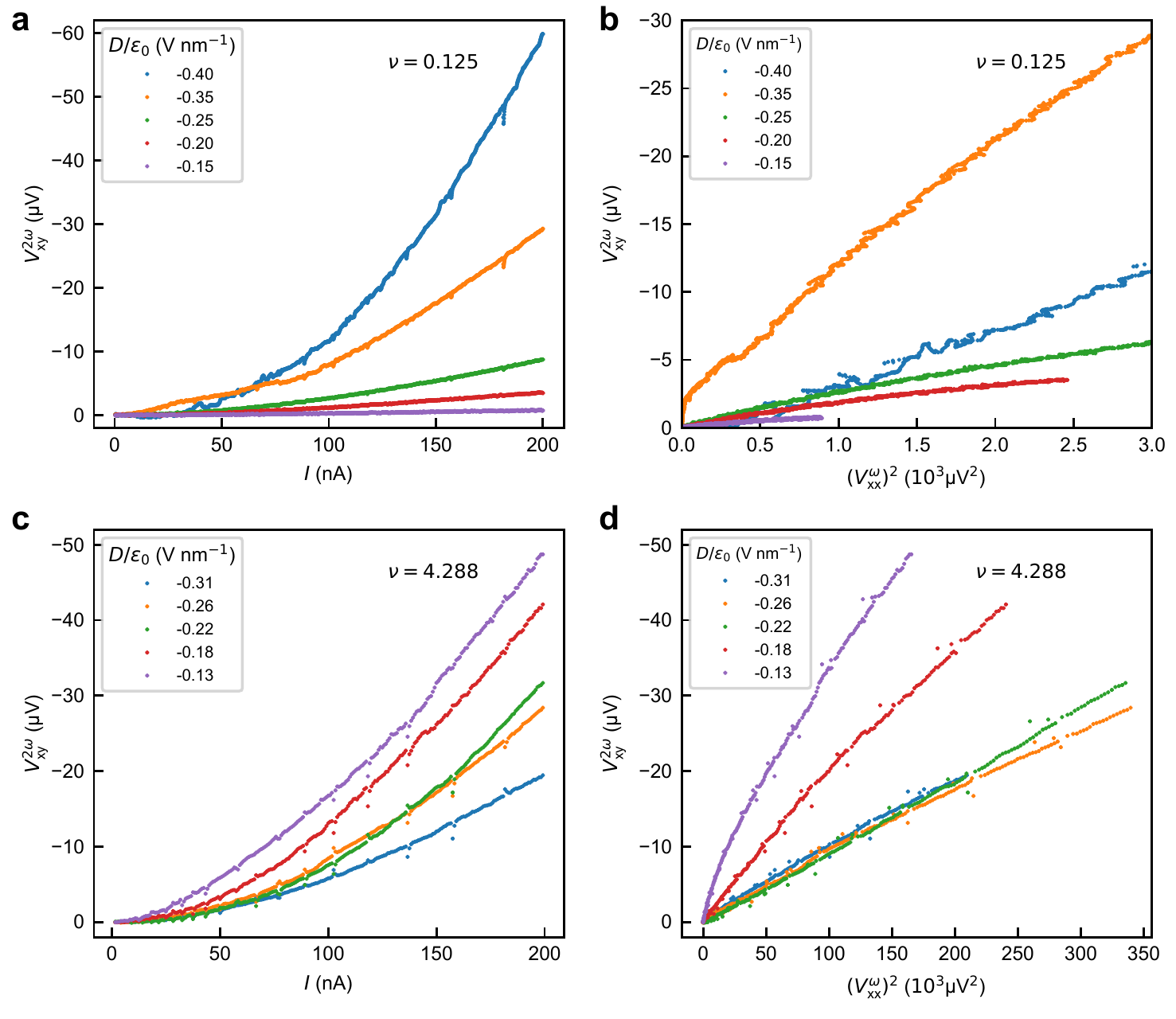}
	\caption{ \label{fig:quadratic} {\footnotesize \textbf{Additional quadratic scaling data.}}
		\textbf{a, b,}~Quadratic dependence of \Vxytw{} on current ($I$) (\textbf{a}) and linear dependence of \Vxytw{} on (\Vxxw)$^2$ (\textbf{b}) for the same filling $\nu=0.125$ as in Fig.~3b of main manuscript, at different displacement fields. \textbf{c, d,}~Quadratic dependence of \Vxytw{} on current ($I$) (\textbf{c}) and linear dependence of \Vxytw{} on (\Vxxw)$^2$ (\textbf{d}) for a filling close to electron-side \moire gap at $\nu=4.288$, for different displacement fields.}
\end{figure*}

\subsection{Phase}
In Supplementary Fig.~\ref{fig:phase}, we show the phase of the measured NLH voltage as a function of displacement field for the filling $\nu=0.125$. 
The filling and the range of displacement field is same to that we explore in Fig.~3a and 3b of main manuscript.
When an ac current $I=I_0 \sin \omega t$ is sent, quadractic scaling of the second harmonic NLH voltage dictates that \Vxytw{} $\propto$ $I^2$ $\propto$ $I_0^2 \sin^2 \omega t=I_0^2(1+\sin(2\omega t-\pi/2))/2$.
The phase remains close to 90$^\circ$ across the two regimes, which is consistent to the second order nature of \Vxytw. 
We additionally note that we measured the out-of-phase component of the NLH voltage consistently throughout all the data presented in main manuscript and supplementary.
\begin{figure*}
	\centering
	\includegraphics[width=10cm]{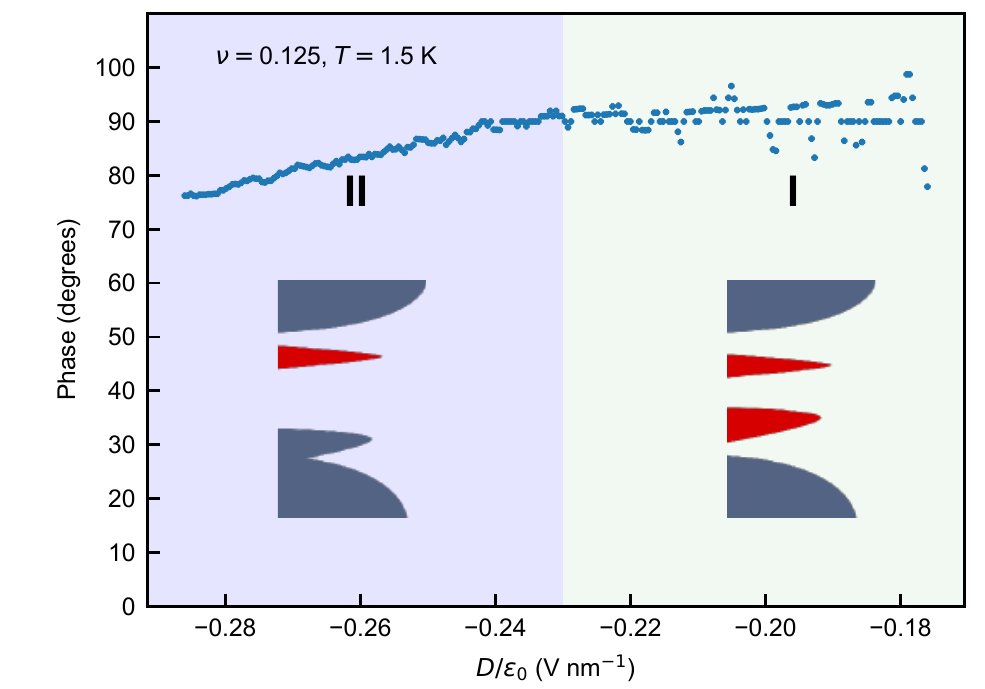}
	\caption{ \label{fig:phase} {\footnotesize \textbf{Phase of nonlinear Hall voltage.}}
		Variation of the phase of the nonlinear Hall voltage with displacement field for the same filling $\nu=0.125$ and temperature $T=1.5$~K as in Fig.~3b of main manuscript. The cartoon indicates the density of states in the two regimes I (light green background) and II (light blue background).}
\end{figure*}

\subsection{NLH voltage at other frequency}
\begin{figure*}
	\centering
	\includegraphics[width=10cm]{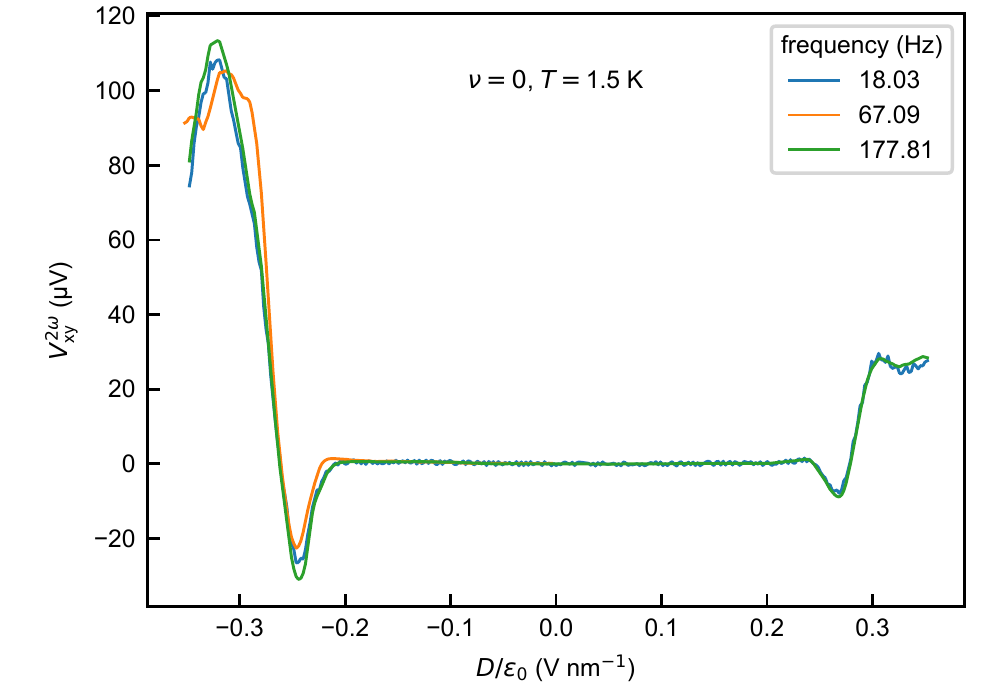}
	\caption{ \label{fig:freq} {\footnotesize \textbf{Frequency dependence of nonlinear Hall voltage.}}
		The nonlinear Hall voltage \Vxytw{} vs. displacement field, measured at a frequency of 2$\omega$ for current with three different frequencies, $\omega =18.03$~Hz (blue line), 67.09~Hz (orange line) and 177.81~Hz (green line). \Vxytw{} remains constant even when the frequency ($\omega$) of current is varied by an order of magnitude. The data is taken for a fixed filling $\nu=0$ and temperature $T=1.5$~K. The data corresponding to orange curve was taken for $D<0$. }
\end{figure*}

Supplementary Fig.~\ref{fig:freq} shows the nonlinear Hall voltage (\Vxytw) dependence on displacement field for three different frequencies of the driving current.
No frequency dependence of \Vxytw{} is observed even when the frequency we explored is varied by an order of magnitude (from $\sim$ 18-178 Hz).
This is consistent to earlier reports~\cite{kang_nonlinear_2019,ma_observation_2019,huang_giant_2021}.
Theoretically, independence of NLH voltage with frequency in the low-frequency regime such that $\omega \tau \to 0$, where $\omega$ is the frequency and $\tau$ is the scattering time, is a signature that the NLH voltage is BCD-induced~\cite{fu_quantum_2015}.

\section{Scaling of NLH voltage with conductivity}

\subsection{\Vxytw/(\Vxxw)$^2$ vs $\sigma_{xx}^2$ using displacement field as parameter for different fixed temperature}

In Supplementary Fig.~\ref{fig:Dfullrange}a, we show the dependence of \Vxytw/(\Vxxw)$^2$ on ($\sigma_{xx}^2$) using $D/\epsilon_0$ as a parameter for the same filling $\nu=0.125$ where the range of $|D|/\epsilon_0$ extends below 0.175~V/nm (in Fig.~3b of main manuscript, $|D|/\epsilon_0$ is varied till 0.175~V/nm).
We see that at lower displacement fields, \Vxytw/(\Vxxw)$^2$ is close to zero.
Interestingly, from Extended Data Fig.~3b, we note that as the magnitude of displacement field is decreased to below 0.18 V/nm, the $R$ vs $T$ at $\nu=0$ becomes metallic, indicating a closing of the gap between the flat bands.
Supplementary Fig.~\ref{fig:Dfullrange}b shows \Vxytw/(\Vxxw)$^2$ dependence on $\sigma_{xx}^2$ at an elevated temperature of $T=12$~K for the same filling $\nu=0.125$. 
Here, we see that the intercept is very close to zero in regime-I as well, that was otherwise nonzero at $T=1.5$~K in Supplementary Fig.~\ref{fig:Dfullrange}a.
The inset in Supplementary Fig.~\ref{fig:Dfullrange}b shows the variation of the y-intercept in regime-I, as a function of temperature.
Systematic variation of the intercept (or, \Vxytw/(\Vxxw)$^2$ for $\sigma \to 0$) with $T$ confirms additionally that \Vxytw/(\Vxxw)$^2$ for $\sigma \to 0$, as we extract in Fig.~3d of main manuscript, is related to BCD. 

\begin{figure*}
	\centering
	\includegraphics[width=16cm]{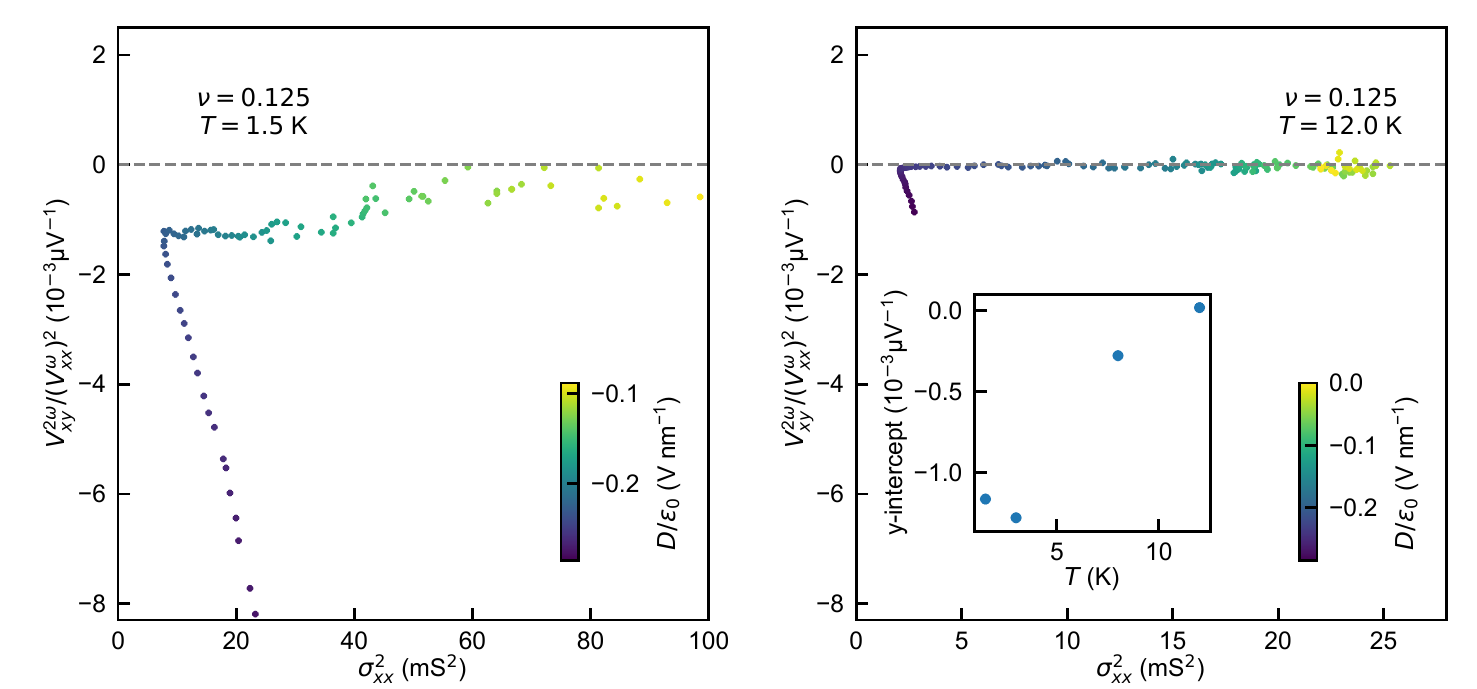}
	\caption{ \label{fig:Dfullrange} {\footnotesize \textbf{Scaling of \Vxytw/(\Vxxw)$^2$ vs $\sigma_{xx}^2$ at elevated temperatures.}}
		\textbf{a, b,}~The variation of normalized nonlinear Hall voltage \Vxytw/(\Vxxw)$^2$ with square of longitudinal conductivity $\sigma_{xx}^2$ plotted parametrically as a function of the displacement field $D/\epsilon_0$ for $T=1.5$~K (\textbf{a}) and $T=12$~K (\textbf{b}) at $\nu=0.125$. The color of the data points indicate the corresponding displacement field. The horizontal dashed gray line at \Vxytw/(\Vxxw)$^2=0$ is a guide to the eye. The inset in \textbf{b} shows the extracted y-intercept of the normalized nonlinear Hall voltage \Vxytw/(\Vxxw)$^2$ for regime-I at few other temperatures for the same filling.}
\end{figure*}

\subsection{Extracting BCD via a second method}
For low frequency, the BCD ($\Lambda$) can also be represented as~\cite{ma_observation_2019,fu_quantum_2015}
\begin{equation}\label{PabloBCDeq}
	\Lambda=\frac{2\hbar^2\sigma_{xx}^3 V_{xy}^{2\omega} W}{e^3\tau I^2}.
\end{equation}
Here, $\sigma_{xx}$ is the longitudinal conductivity, \Vxytw{} is the nonlinear Hall voltage, W is the width of TDBG, $\tau$ is the scattering time, $I$ is the current sent with low frequency and $e$ is the electronic charge.
Using Drude formula, $\sigma_{xx}=ne^2\tau/m$, where $n$ is the charge density and m is the effective mass, and $\sigma_{xx}=\frac{I L}{V_{xx}^{\omega} W}=\frac{I}{V_{xx}^{\omega}}$ (for our case, length ($L$)=width ($W$)=2~$\mu m$) in eq.~\eqref{PabloBCDeq}, we obtain
\begin{equation}\label{PabloBCDeq2}
	\Lambda=\frac{2\hbar^2 W n}{e m}\times \left(\frac{V_{xy}^{2\omega}}{(V_{xx}^{\omega})^2}\right).
\end{equation}

In Supplementary Fig.~\ref{fig:PabloBCD}, we plot the BCD using eq.~\eqref{PabloBCDeq2} using $\frac{V_{xy}^{2\omega}}{(V_{xx}^{\omega})^2}$ for $\sigma_{xx}\to0$ (the y-intercept in Fig.~3b of main manuscript) for different fillings $\nu=4n/n_\text{S}$. 
Here, $n_\text{S}=2.8 \cross 10^{12}$~cm$^{-2}$ is the required charge density to completely fill the flat conduction band, corresponding to a twist angle of 1.1$^\circ$. 
With an assumption of m=3$m_e$, where $m_e$ is mass of electron, we get a good agreement with the BCD extracted in Fig.~3c of main manuscript.

\begin{figure*}
	\centering
	\includegraphics[width=15.5cm]{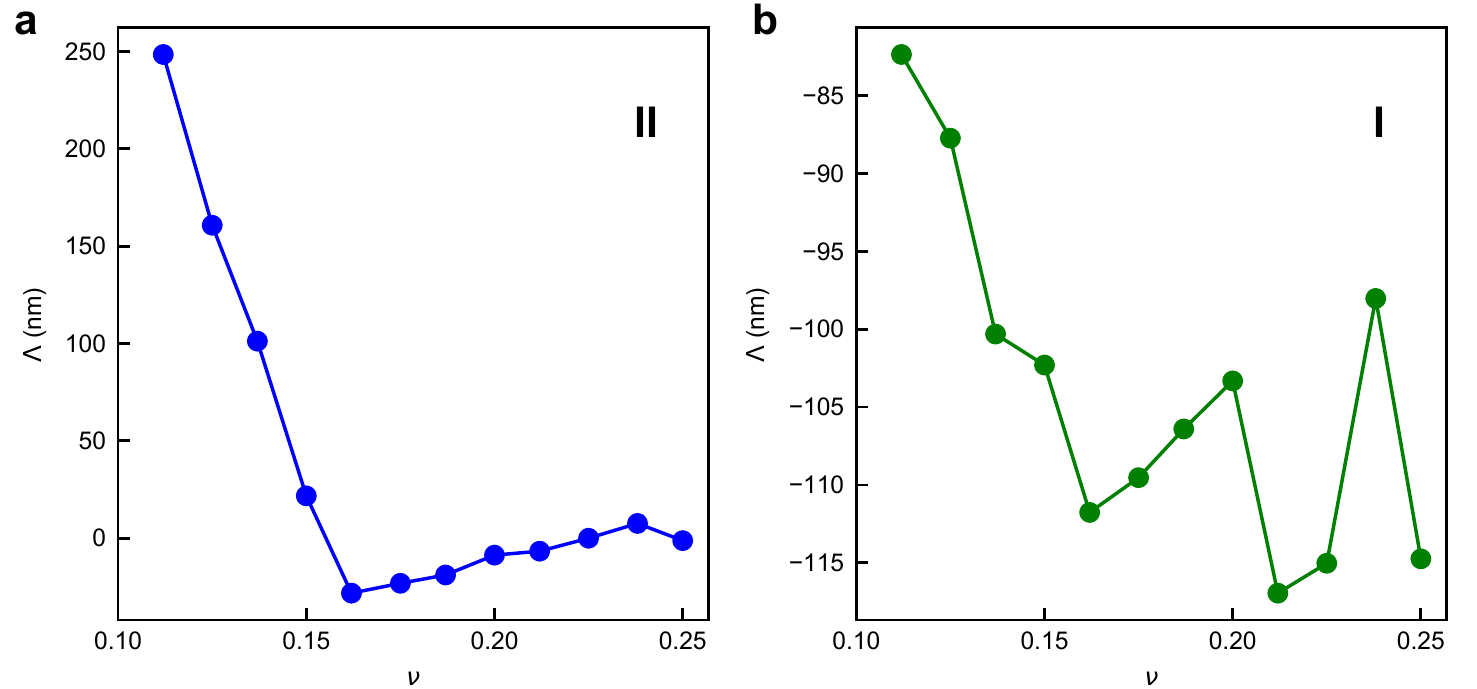}
	\caption{ \label{fig:PabloBCD} {\footnotesize 
			\textbf{a, b,}~Extracting the BCD as a function of filling ($\nu$) using eq.~\eqref{PabloBCDeq2} for regime-II~(\textbf{a}) and regime-I~(\textbf{b}).
	}}
\end{figure*}
\clearpage

\section{Additional hysteresis data}
\subsection{Repeatability and histogram}
\begin{figure*}[!h]
	\centering
	\includegraphics[width=15.5cm]{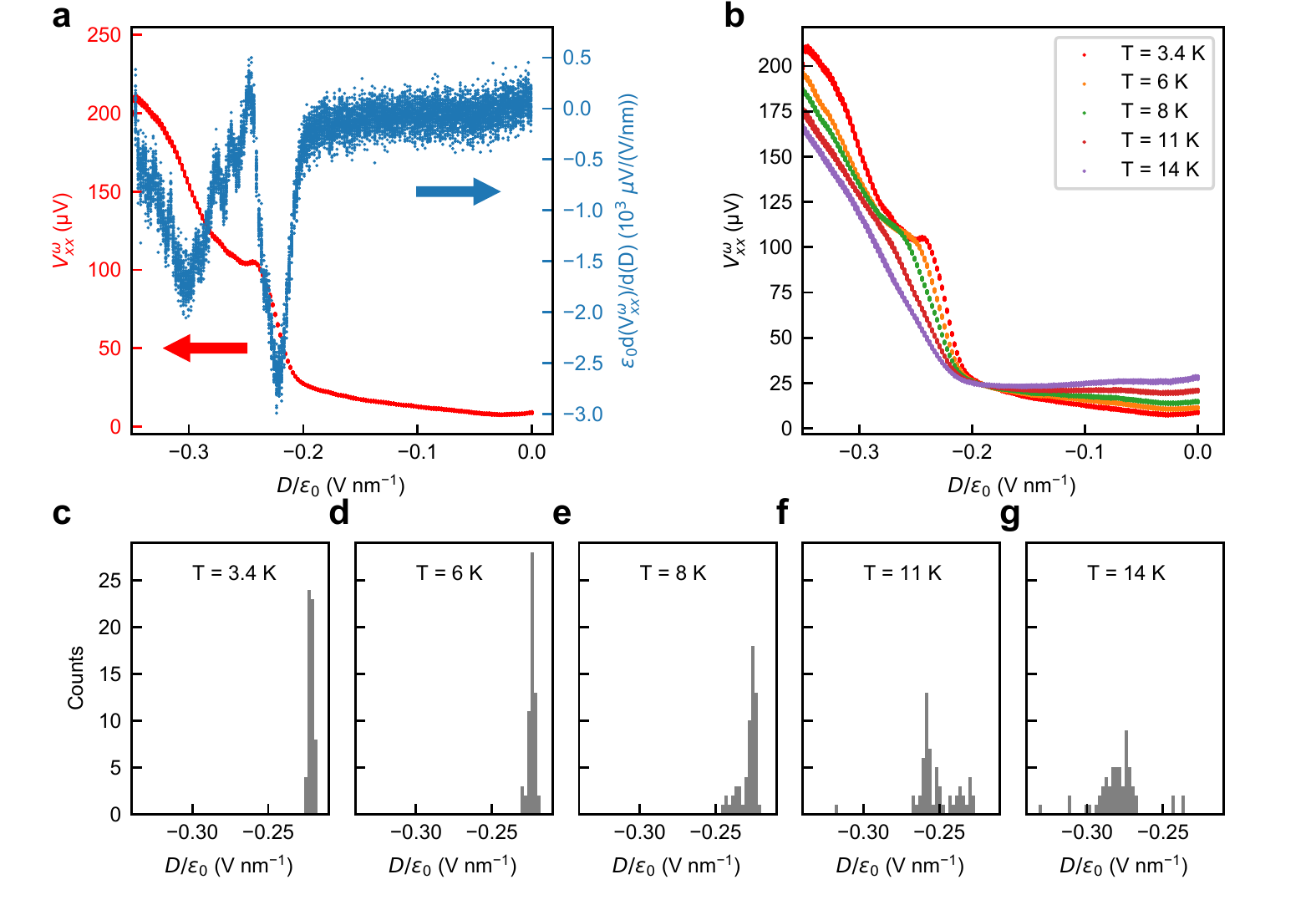}
	\caption{ \label{fig:histogram} {\footnotesize 
			\textbf{Switching statistics as a function of temperature.}}
		\textbf{a,}~Variation of longitudinal voltage \Vxxw{} with displacement field $D/\epsilon_0$ (red colored plot corresponding to left axis, as indicated by the red arrow) when $D/\epsilon_0$ is swept from left to right along the x-axis at a constant rate of 2.5~mV nm$^{-1}$ s$^{-1}$. 
		The blue colored plot (corresponding to the right axis) shows the variation of the numerical derivative $\frac{dV_{xx}^\omega}{d(D/\epsilon_0)}$ with $D/\epsilon_0$.
		The data is plotted for the same direction of $D/\epsilon_0$ for 59 cycles at a constant temperature of $T=3.4$~K, indicating repeatability. 
		\textbf{b,}~The variation of \Vxxw{} with $D/\epsilon_0$ at few fixed temperatures. 
		For each temperature, the data is plotted for 59 up cycles.
		\textbf{c-g,}~Switching statistics at few fixed temperatures as mentioned in each subpanel. The histogram counts the position (in $D/\epsilon_0$ axis) of the minima of $\frac{dV_{xx}^\omega}{d(D/\epsilon_0)}$ for the 59 up cycles at each fixed temperature.}
\end{figure*}
\newpage

\subsection{Temperature dependence}
\begin{figure*}[!h]
	\centering
	\includegraphics[width=15.5cm]{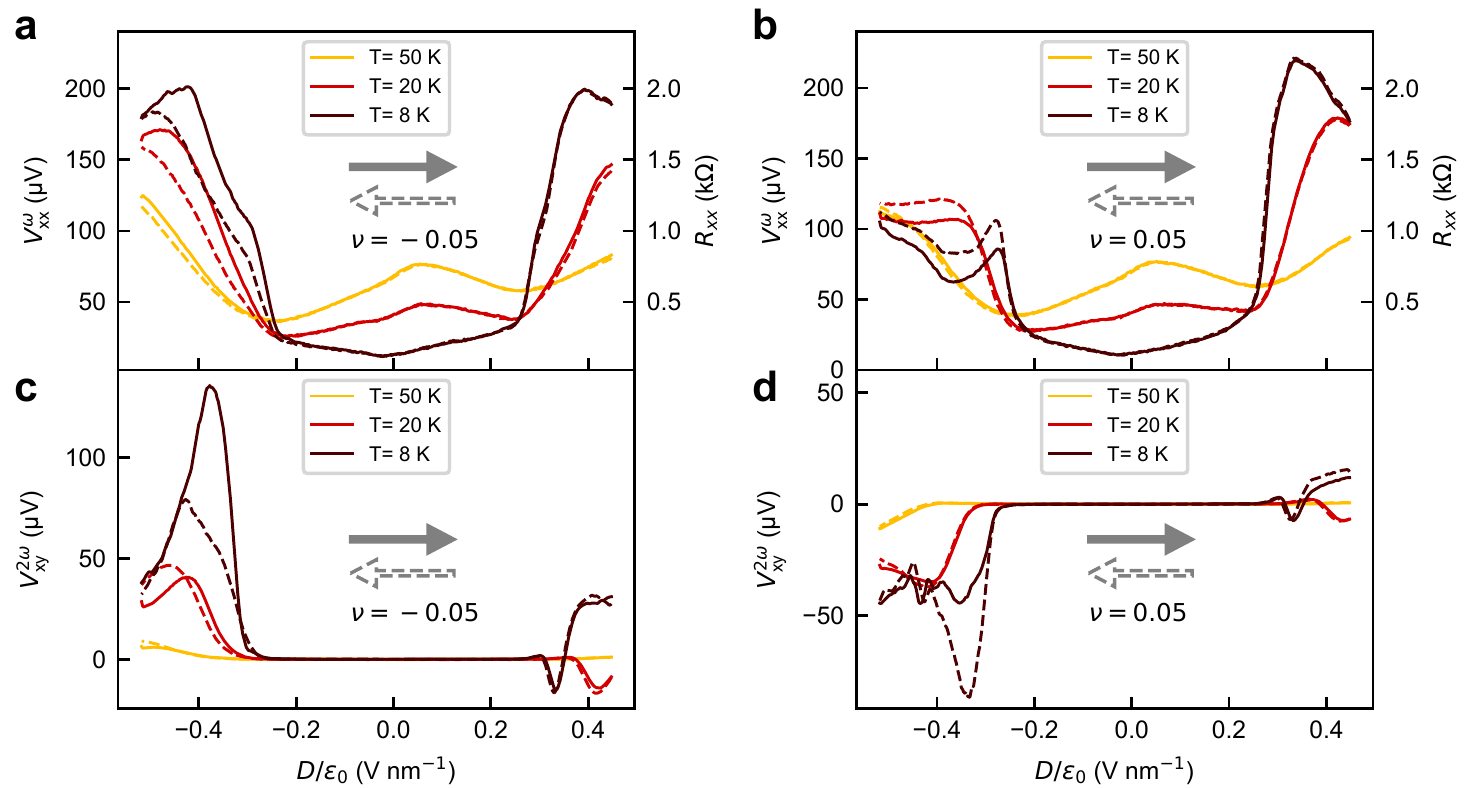}
	\caption{ \label{fig:Dhyst_T} {\footnotesize \textbf{Temperature dependence of hysteresis in longitudinal and nonlinear Hall voltages.} \textbf{a, b,}~Variation of the longitudinal voltage, \Vxxw{} with perpendicular displacement field ($D/\epsilon_0$) for $\nu=-0.05$~(\textbf{a}) and $\nu=0.05$~(\textbf{b}) at different fixed temperatures. The right axes indicate the corresponding longitudinal resistance, $R_{xx}$. \textbf{c, d,}~Variation of the corresponding nonlinear Hall voltage, \Vxytw{} for $\nu=-0.05$~(\textbf{c}) and $\nu=0.05$~(\textbf{d}) with $D/\epsilon_0$. The solid and dashed lines stand for the voltage response with increasing and decreasing values of $D/\epsilon_0$, respectively. The arrows indicate the sweep direction of $D/\epsilon_0$. The displacement field was swept at a constant rate of 2.5~mV nm$^{-1}$ s$^{-1}$.}}
\end{figure*}
\newpage

\subsection{Effect of current and frequency}
\begin{figure*}[!h]
	\centering
	\includegraphics[width=15.5cm]{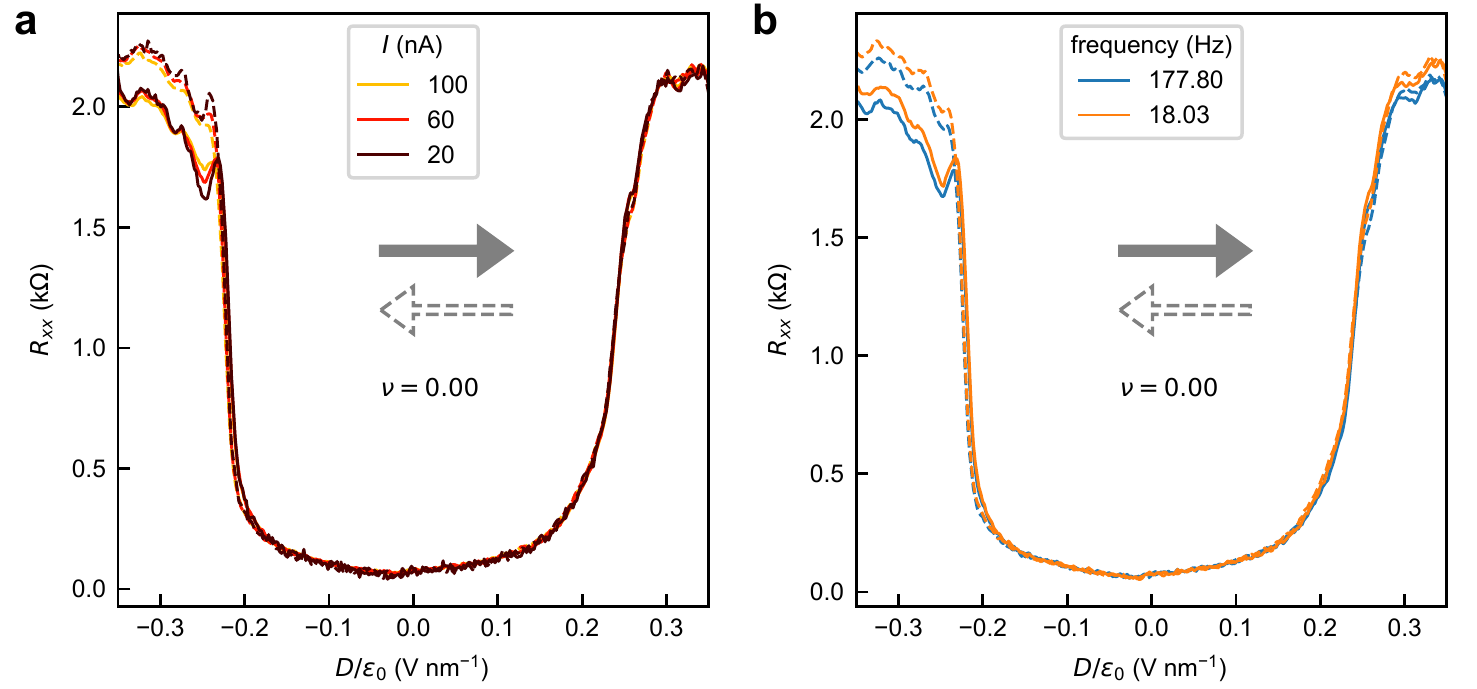}
	\caption{ \label{fig:Hyst_Iandf_effect} {\footnotesize Current ($I$) (\textbf{a}) and frequency (\textbf{b}) dependence of hysteresis in longitudinal resistance $R_{xx}$ with displacement field ($D/\epsilon_0$) at a fixed temperature $T=1.5$~K. The displacement field was swept at a constant rate of 2.5~mV nm$^{-1}$ s$^{-1}$. Independence of hysteresis on value of driving current in \textbf{a} rules out any heating effect. Independence of hysteresis on frequency in \textbf{b} rules out any effect of stray capacitance.}}
\end{figure*}
\newpage

\subsection{Data for other fixed $\nu$}
\begin{figure*}[!h]
	\centering
	\includegraphics[width=15.5cm]{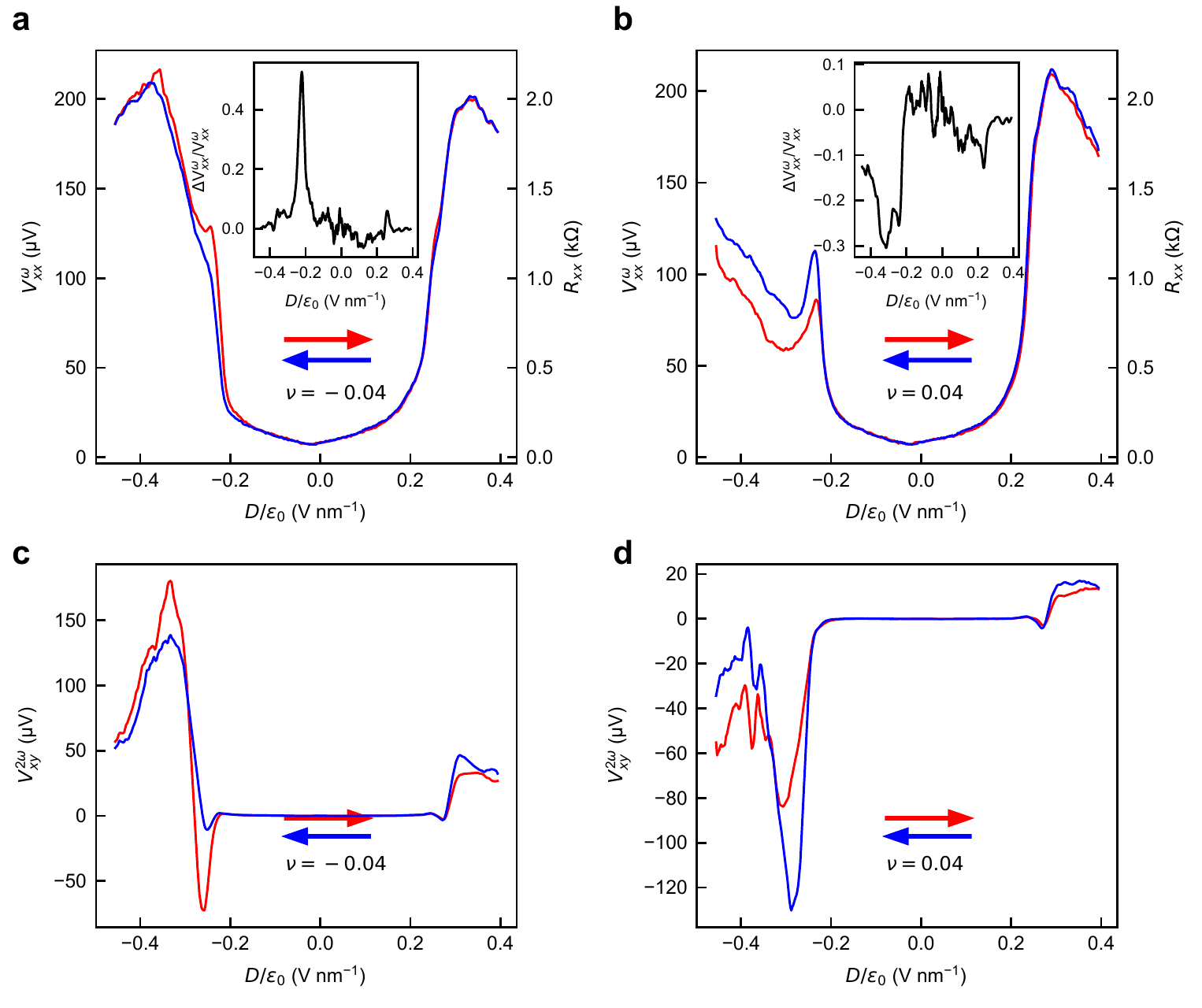}
	\caption{ \label{fig:Hyst_doping} {\footnotesize \textbf{Dependence of sense of hysteresis on doping.~a, b,}~ Hysteresis in longitudinal resistance $R_{xx}$ with displacement field ($D/\epsilon_0$) for a doping below CNP at $\nu=-0.04$ (\textbf{a}) and above CNP at $\nu=0.04$ (\textbf{b}). The arrows indicate direction of sweeping $D$. On the negative $D$ side, the red curve leads the blue curve in \textbf{a}, while it lags in \textbf{b}. 
			Such a change in hysteretic response due to doping cannot be accounted for via charge traps in dielectric.
			The insets show the relative difference in \Vxxw{} of up and down sweep. The relative difference is maximum around $D/\epsilon_0=-0.23$~\Vbynm{}, where a topological transition takes place as discussed in main manuscript.
			Hysteresis in \Vxytw{} with displacement field ($D/\epsilon_0$) for a doping below CNP at $\nu=-0.04$ (\textbf{c}) and above CNP at $\nu=0.04$ (\textbf{d}), showing similar flipping in sense of hysteresis.
			The temperature was fixed at $T=1.5$~K and the displacement field was swept at a constant rate of 2.5~mV nm$^{-1}$ s$^{-1}$.}}
\end{figure*}
\newpage

\subsection{Rate dependence}
\begin{figure*}[!h]
	\centering
	\includegraphics[width=16cm]{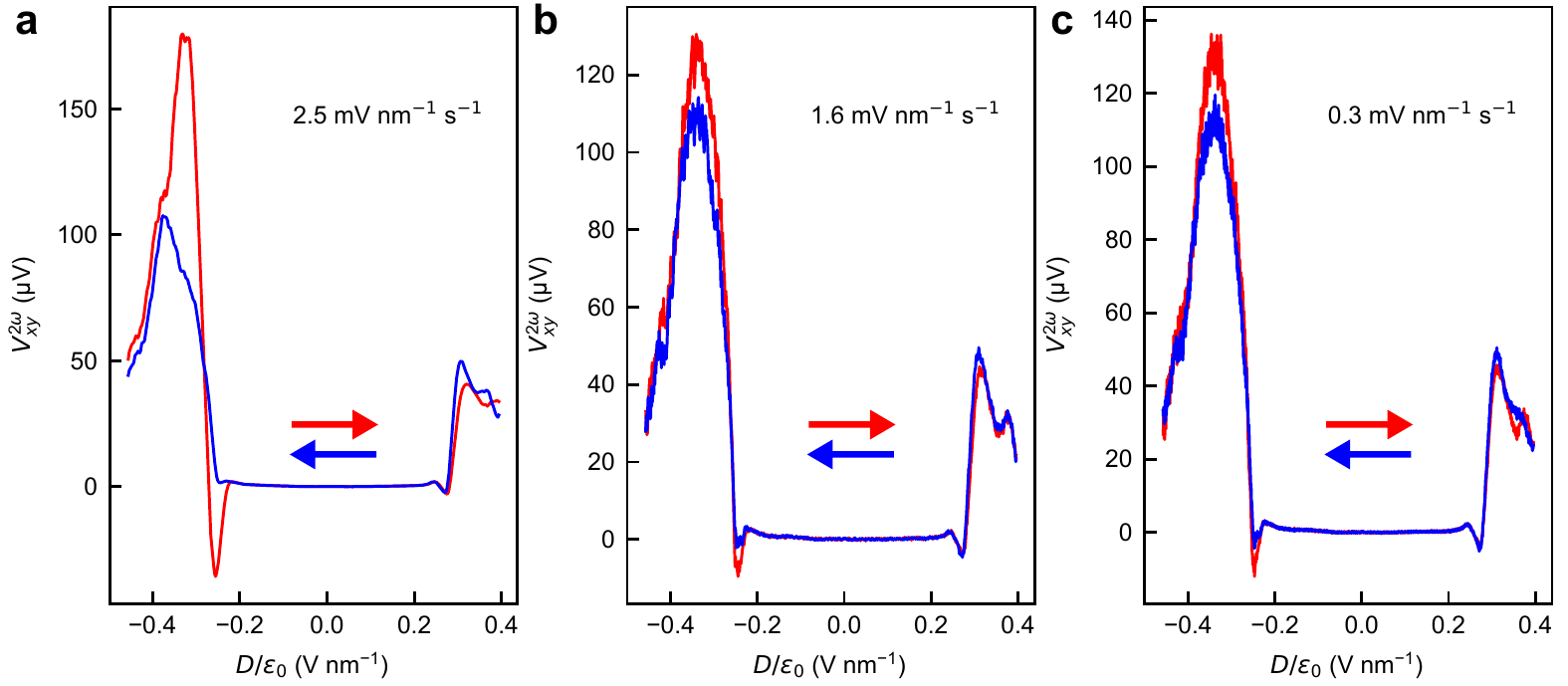}
	\caption{ \label{fig:Hyst_rate} {\footnotesize \Vxytw{} vs $D/\epsilon_0$ hysteresis at a fixed $\nu=-0.05$ for three different rates of sweeping $D/\epsilon_0$ at 2.5 mV nm$^{-1}$s$^{-1}$ (\textbf{a}), 1.6 mV nm$^{-1}$s$^{-1}$ (\textbf{b}), and 0.3 mV nm$^{-1}$s$^{-1}$ (\textbf{c}). }}
\end{figure*}
\newpage

\subsection{hysteresis data from another device}

\begin{figure*}[!h]
	\centering
	\includegraphics[width=15.5cm]{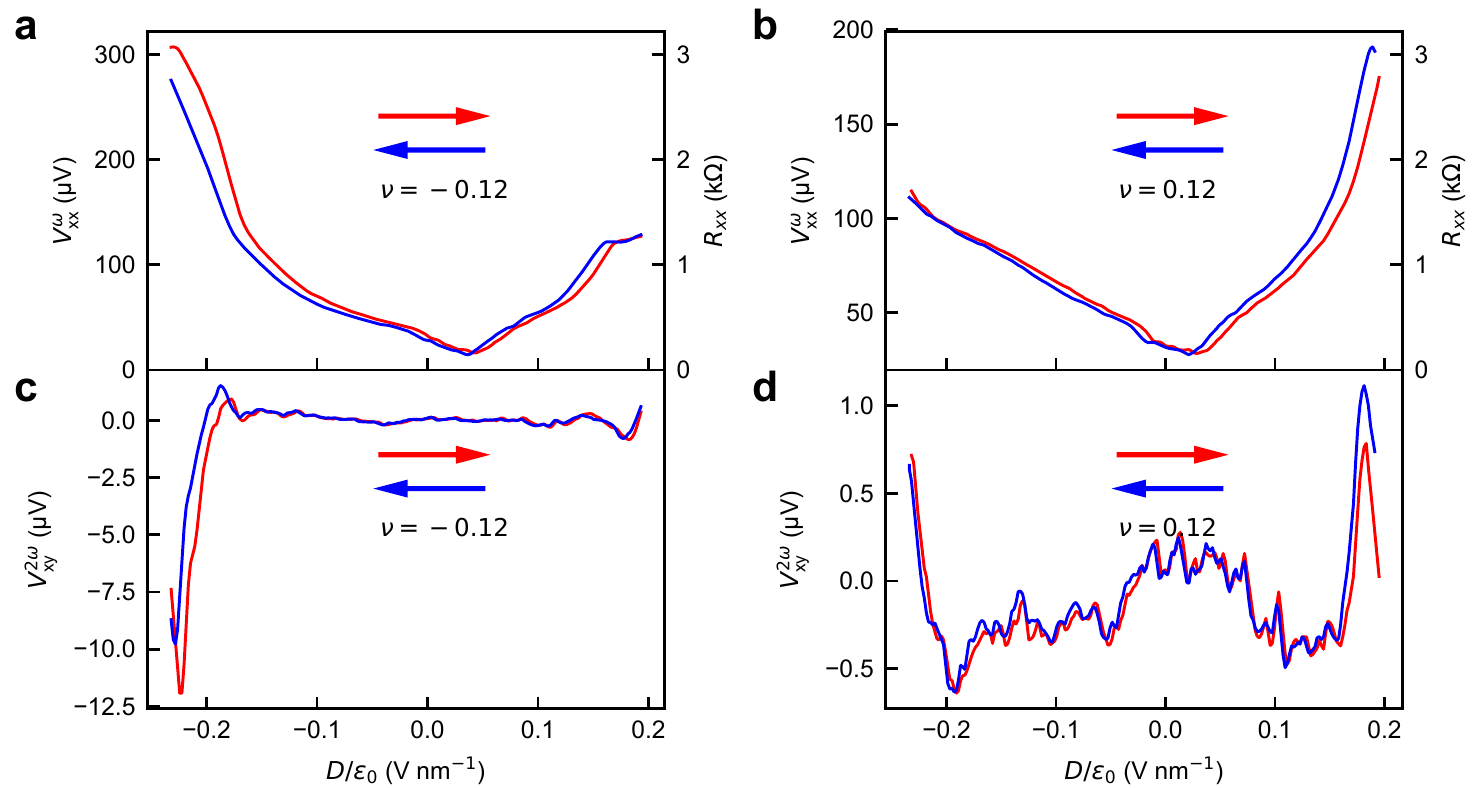}
	\caption{ \label{fig:hyst_second_dev} {\footnotesize 
			\textbf{Hysteresis in device-2 with twist angle of 1.26$^\circ$.}}
		\textbf{a, b,}~Variation of the longitudinal voltage \Vxxw{} with perpendicular displacement field ($D/\epsilon_0$) for a small negative filling~(\textbf{a}) and positive filling~(\textbf{b}) around CNP ($\nu=0$). The right axes indicate the corresponding longitudinal resistance, $R_{xx}$. \textbf{c, d,}~Variation of the nonlinear Hall voltage \Vxytw{} with $D/\epsilon_0$ for similar small negative doping~(\textbf{c}) and positive doping~(\textbf{d}) around CNP. The red and blue solid lines stand for the voltage response with increasing and decreasing values of $D/\epsilon_0$, respectively. The arrows indicate the sweep direction of $D/\epsilon_0$. The displacement field was swept at a constant rate of 2.7~mV nm$^{-1}$ s$^{-1}$. The temperature was fixed at 10~mK. Measurement was performed with a current of 100~nA at a frequency of 177~Hz.}
\end{figure*}
\newpage

\section{Polarization calculation details}
\subsection{Computational Details}
Our first-principles calculations are based on density functional theory (DFT) as implemented in Quantum ESPRESSO (QE)  package \cite{giannozzi2009quantum}. We use ultrasoft pseudopotentials to represent the interaction between ionic cores and valence electrons. The exchange-correlation energy of electrons is treated within a generalized gradient approximation (GGA) \cite{hua1997generalized} with a functional form parameterized by Perdew, Burke, and Ernzerhof \cite{perdew1998perdew}. We truncated the plane wave basis used in expansion of Kohn-Sham wave functions and charge density with energy cut-offs of 45 Ry and 360 Ry respectively. We used $18\times18\times1$ uniform grid of k-points for sampling the Brillouin zone (BZ) integrations. The discontinuity in occupation numbers of electronic states was smeared with broadening temperature of \textit{k}\textsubscript{B}\textit{T} = 0.005 Ry using a Fermi-Dirac distribution function. We include van der Waals (vdW) interaction using PBE + D2 method of Grimme \cite{grimme2004accurate}. The 2D system is simulated using a periodic supercell, with a vacuum layer of 12 {\AA} separating adjacent periodic images of the sheet. To simulate response to electric field, we add a saw-tooth potential as a function \textit{z}.

\subsection{Results and Discussion}
\begin{figure*}
	\centering
	\includegraphics[width=15.5cm]{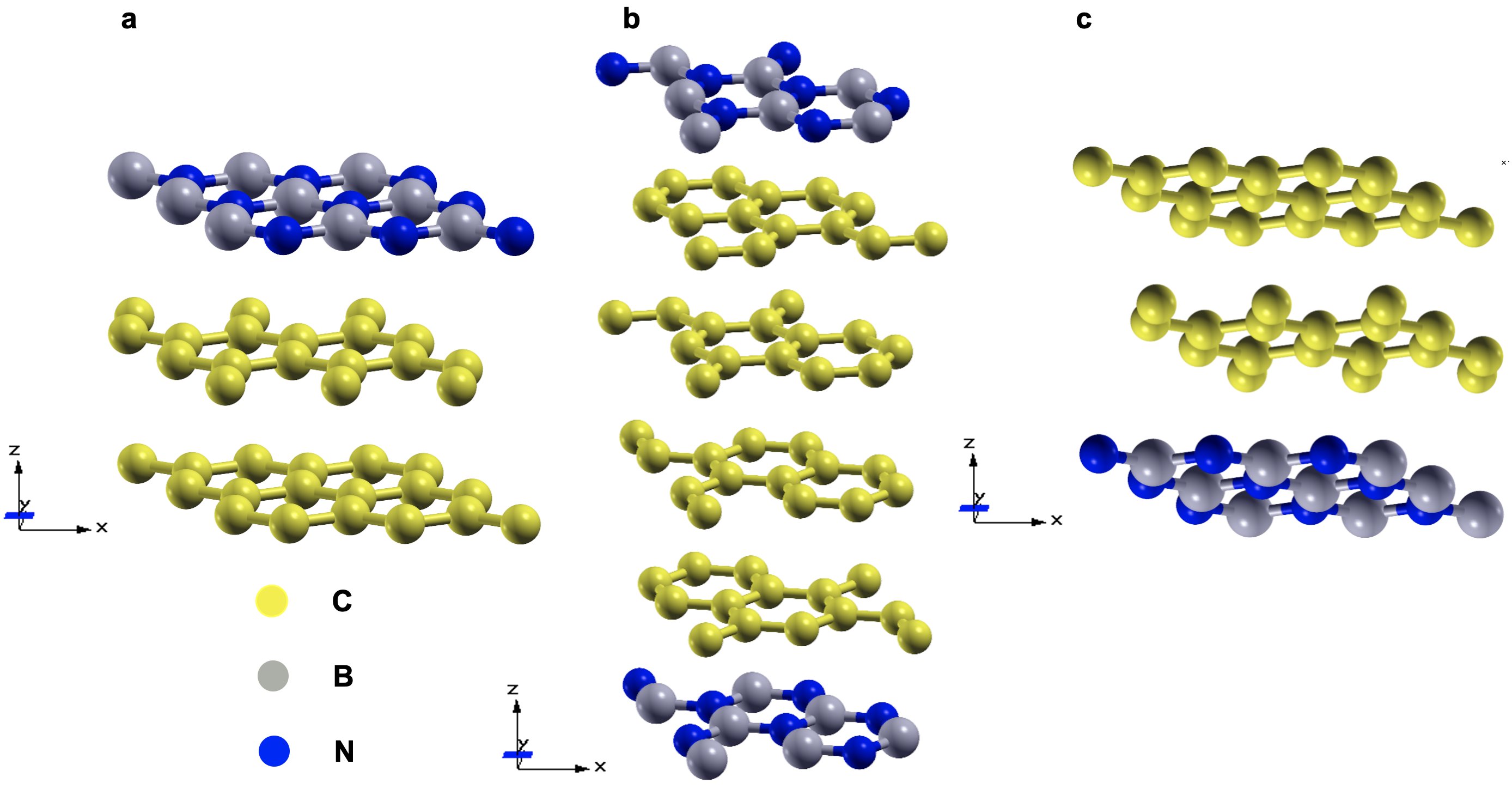}
	\caption{\footnotesize  \textbf{a,}~Structure of Gr-Gr-h-BN.
		\textbf{b,}~Unit cell of twisted double bilayer graphene encapsulated between hexagonal boron nitride (h-BN-TDBG-h-BN) with a rotation angle of 21.78\textsuperscript{o}.
		\textbf{c,}~Structure of h-BN-Gr-Gr. 
		The building blocks of h-BN-TDBG-h-BN are Gr-Gr-h-BN and h-BN-Gr-Gr. The atomic species C, B and N are displayed in yellow, grey and blue colors, respectively.}
	\label{fig:Sfig_polarization1}
\end{figure*}

Twisted double bilayer graphene, TDBG, in our experiments is composed of two AB-stacked bilayers of graphene rotated by an angle~\cite{PhysRevB.88.035428} $\theta$ and encapsulated from top and bottom by hexagonal boron nitride (h-BN-TDBG-h-BN) (Supplementary Fig.~\ref{fig:Sfig_polarization1}b).
Thus, parallelly stacked one atomic plane of hexagonal boron nitride and two layers of graphene (Gr-Gr-h-BN) is a building block making a half of h-BN-TDBG-h-BN (upper trilayer) (Supplementary Fig.~\ref{fig:Sfig_polarization1}a). 
Interestingly, a single Gr-Gr-h-BN unit is noncentrosymmetric and lacks the horizontal mirror symmetry. 
Hence, it is expected to have a non-vanishing polarization (dipole moment perpendicular to the plane).
We consider two different configurations of Gr-Gr-h-BN which have been obtained by changing stacking sequences of Gr-Gr-h-BN (i) stacking of h-BN same as bottom graphene and (ii) stacking sequence of h-BN not matching with either of the two graphene layers.
However, the energies of the two configurations are comparable (0.1 meV) and we carry out our theoretical analysis with the first configuration as a model. 

\begin{figure*}
	\centering
	\includegraphics[width=15.5cm]{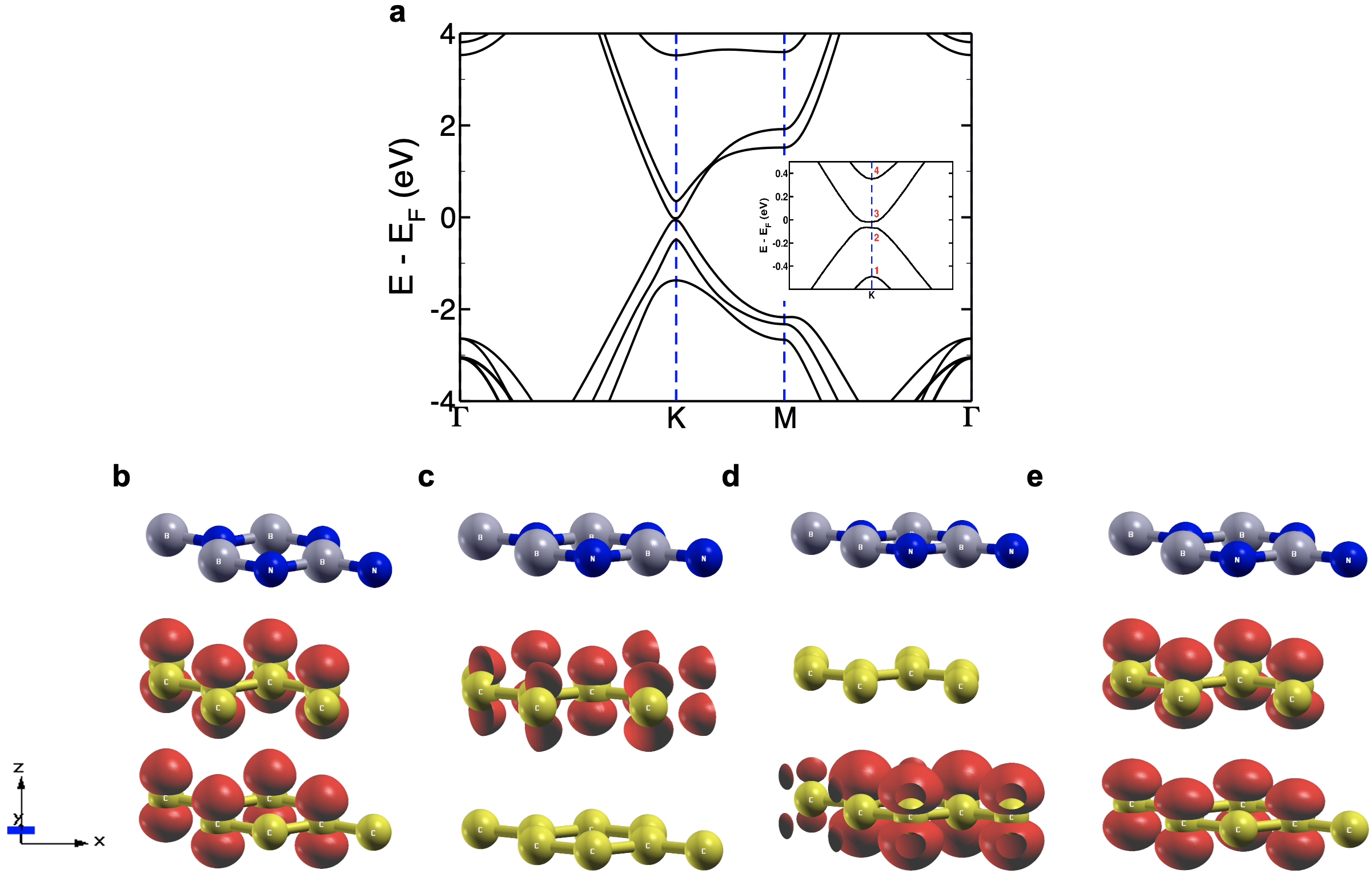}
	\caption{\footnotesize \textbf{a,}~Electronic structure of Gr-Gr-h-BN shows a band gap of 26 meV at \textit{K} point. 
		\textbf{b-e,}~Visualization of wavefunctions of four states near Fermi energy at \textit{K} point of Band 1~(\textbf{b}), Band 2~(\textbf{c}), Band 3~(\textbf{d}) and Band 4~(\textbf{e}) of Gr-Gr-h-BN shows contribution from \textit{p}\textsubscript{z} orbitals of carbon of graphene. The average position, $\langle{Z}\rangle$ in terms of interlayer distance d = 3.2 {\AA} for bands 2 and 3 is 3/2d,  and 1/2d, respectively.}
	\label{fig:Sfig_polarization2}
\end{figure*}

We first examine the electronic structure of Gr-Gr-h-BN and find a band gap of $\sim$ 26 meV at \textit{K} point (h-BN breaks sublattice symmetry of AB-stacked bilayer graphene, Supplementary Fig.~\ref{fig:Sfig_polarization2}a). 
From the slope of macroscopic average electrostatic potential in vacuum, our estimate of polarization of Gr-Gr-h-BN is \textit{P}\textsubscript{z} $\approx$ -0.34 $\mu$C/cm\textsuperscript{2} (Extended Data Fig. 5b). 
Thus, Gr-Gr-h-BN has a nonzero polarization due to the broken inversion and horizontal reflection symmetries. 
This mechanism is similar to ferroelectricity in bilayer h-BN~\cite{Yasuda1458}. 
The lower trilayer of TDBG is h-BN-Gr-Gr~(Supplementary Fig.~\ref{fig:Sfig_polarization1}c) and has exactly the same polarization with opposite sign. 

The sense of hysteresis in longitudinal and nonlinear Hall voltage for doping just below charge neutrality point (CNP) is flipped on changing the doping to a point just above CNP. 
To understand the metastable states governing this hysteresis, we present a rigid band model for h-BN-Gr-Gr (lower trilayer) and Gr-Gr-h-BN (upper trilayer) of electronic states coupling with electric field and determine polarization as a function of electric field. Evolution of band energies with perpendicular electric field is modeled as
\begin{equation}
	H\textsubscript{\textit{i}} = \epsilon\textsubscript{\textit{i}} + e\textit{E} \langle{Z\textsubscript{\textit{i}}}\rangle
\end{equation} 
where \textit{i} is the band index (\textit{i} = 1 to 4 for the four bands close to Fermi at \textit{K} point), $\epsilon$\textsubscript{\textit{i}} is energy of \textit{i}\textsuperscript{th} band, \textit{e} is charge of an electron, \textit{E} is electric field, $\langle{Z}\rangle$ is the average position of each state~(Supplementary Fig.~\ref{fig:Sfig_polarization2}b-e). 
$\langle{Z}\rangle$ is minus (-) for h-BN-Gr-Gr (lower trilayer) and plus (+) for Gr-Gr-h-BN (upper trilayer), respectively. 
Band 2 of the lower trilayer and band 3 of upper trilayer cross at E = -0.0039 V/{\AA} (band 3 of lower trilayer and band 2 of upper trilayer cross at E = 0.0039 V/{\AA}), resulting in redistribution of charges among these bands (Extended Data Fig.~5c). 
We show that there exist two metastable states for (i) E $<$ -0.0039 V/{\AA} and (ii) E $>$ 0.0039 V/{\AA}, with distinct polarization states in h-BN-TDBG-h-BN, that can be accessed with electric field within our rigid band model (Extended Data Fig.~5d). 
However, the resistance associated with the two metastable states remains the same. 
While the metastable states in h-BN-TDBG-h-BN are explained using the rigid band model, the hysteresis in resistance seen in experiments can originate from the broken symmetry, which can possibly arise from the distinction in coupling of top and bottom gates inducing inhomogeneous doping in the channel or by a twist between two trilayers.

\begin{figure*}
	\centering
	\includegraphics[width=15.5cm]{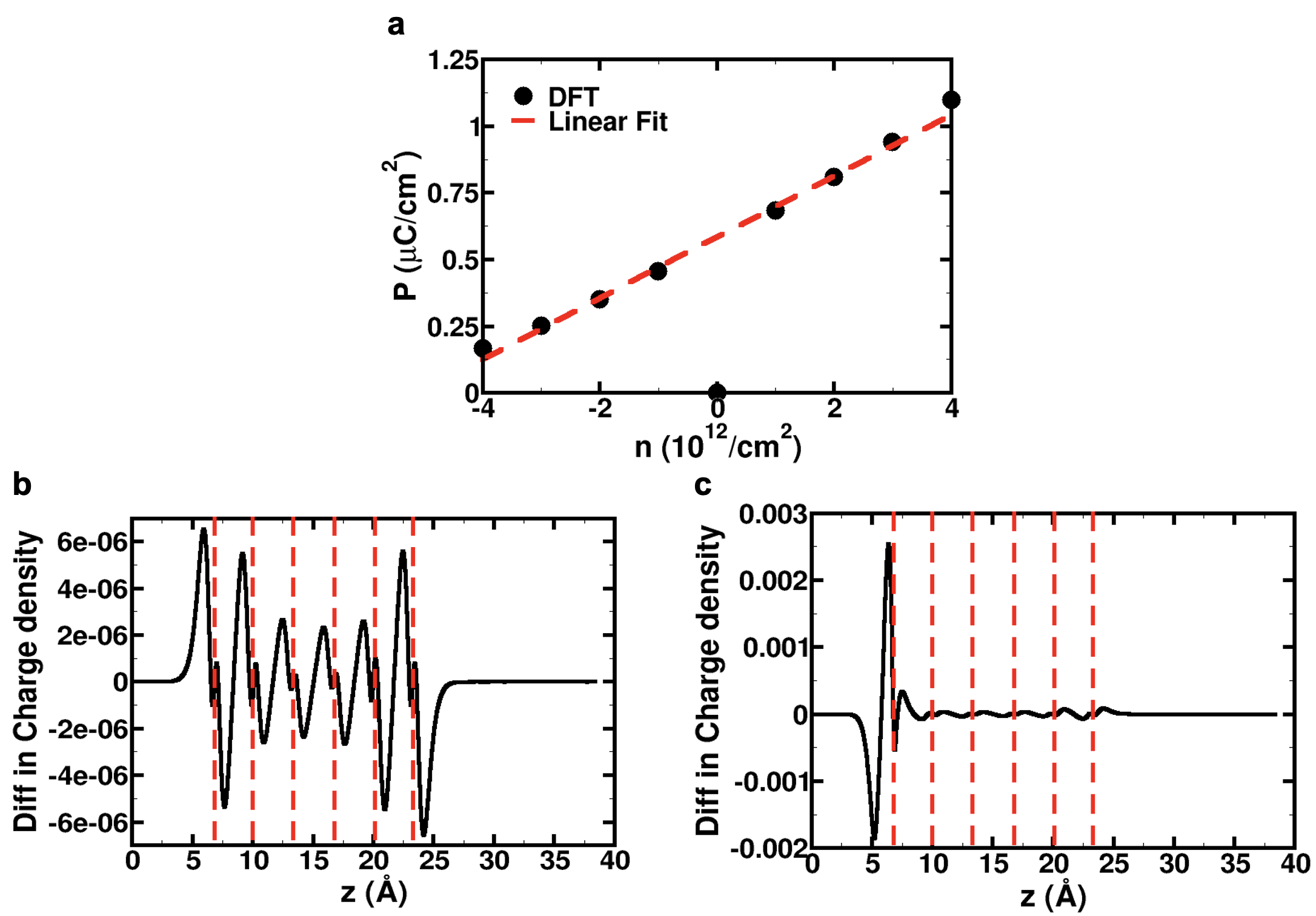}
	\caption{\footnotesize \textbf{a,}~Polarization in TDBG calculated as a function of doping in the presence of a bottom gate. 
		\textbf{b, c,}~The difference in planar-averaged electron charge density, $\bar{\rho}$(\textit{z}) for \textit{n} = 4 $\times$10\textsuperscript{12}/cm\textsuperscript{2}, E = 0.00625 V/{\AA} and \textit{n} = 4 $\times$10\textsuperscript{12}/cm\textsuperscript{2}, E = 0 V/{\AA} without (\textbf{b}) and with (\textbf{c}) gate set-up in h-BN-TDBG-h-BN with a twist angle $\theta$ = 21.78\textsuperscript{o}. In the presence of gate, a positively oriented electric field pushes the electrons at the bottom gate (the scale of y-axis in \textbf{c} is higher than \textbf{b}).}
	\label{fig:Sfig_polarization3}
\end{figure*}

To understand the role of a gate electrode, we obtain the difference in planar-averaged electron charge density, $\bar{\rho}$(\textit{z}) for \textit{n} = 4 $\times$10\textsuperscript{12}/cm\textsuperscript{2}, E = 0.00625 V/{\AA} and \textit{n} = 4 $\times$10\textsuperscript{12}/cm\textsuperscript{2}, E = 0 V/{\AA} with and without electric boundary conditions of a gate (Supplementary Fig.~\ref{fig:Sfig_polarization3}b and~\ref{fig:Sfig_polarization3}c) in h-BN-TDBG-h-BN with a twist angle $\theta$ = 21.78\textsuperscript{o}. 
Asymmetry in $\Delta\bar{\rho}(\textit{z})$ at the atomic planes (red dashed lines in Supplementary Fig.~\ref{fig:Sfig_polarization3}b and~\ref{fig:Sfig_polarization3}c) indicate accumulation and depletion of electronic charge and local polarization arising from polarizability of \textit{p}\textsubscript{z} orbitals and a transfer of a tiny amount of charge. 
In the presence of gate, a positive electric field (along $\hat{z}$) pushes the electrons to the bottom gate (the scale of y-axis in Supplementary Fig.~\ref{fig:Sfig_polarization3}c is higher than Supplementary Fig.~\ref{fig:Sfig_polarization3}b), highlighting the inhomogeneity in doping. 
We note that a spontaneous electric dipole \textit{p}\textsubscript{z} can also arise from the restructuring of the regions with AA, AB, BA and BB stacking upon application of electric field~\cite{Yasuda1458} and contribute to the observed hysteresis.

\section{DC voltage}
Apart from the second harmonic voltage \Vxytw, the nonlinear Hall effect also gives rise to a DC voltage~\cite{fu_quantum_2015,kang_nonlinear_2019}. 
In Supplementary Fig.~\ref{fig:DC}, we show the measured DC voltage in perpendicular direction to an ac current applied with frequency 177 Hz.
The dependence of the DC voltage is shown for the full parameter space of ($\nu$,$D$) for the same device used to show the NLH voltage dependence in Fig.~2b of main manuscript.
DC voltage, together with the measured second harmonic \Vxytw, provides additional evidence for NLH effect in TDBG devices.
\begin{figure*}
	\centering
	\includegraphics[width=8cm]{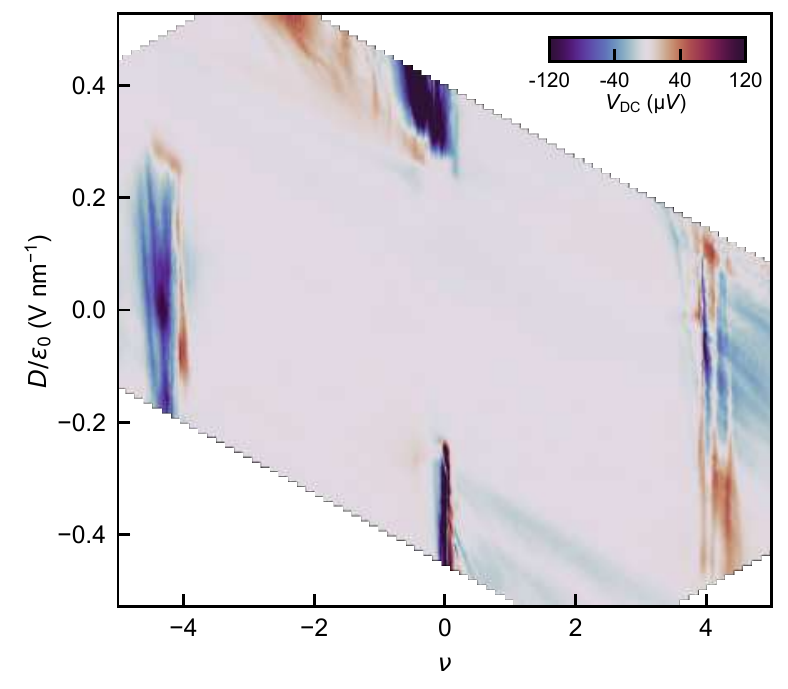}
	\caption{ \label{fig:DC} {\footnotesize 
			\textbf{Dependence of DC voltage on filling factor and displacement field.}}
		DC voltage ($V_{DC}$) as a function of filling factor ($\nu$) and displacement field ($D/\epsilon_0$) measured perpendicular to current $I=100$~nA applied with a frequency of 177~Hz. The color bar provided in top right indicates the corresponding values of $V_{DC}$.
	}
\end{figure*}

\newpage

\end{document}